\documentclass[a4paper,11pt]{article}
\usepackage{jheppub}
\usepackage[T1]{fontenc}
\usepackage{lmodern}
\usepackage{amsmath,amssymb}
\usepackage{xcolor}
\newcommand{\dd}{\mathrm{d}}
\newcommand{\Z}{\mathbb Z}
\title{Second-order localization and intersecting O6-planes in scale-separated AdS vacua}
\author{Aravind Aikot}
\author{and Timm Wrase}
\affiliation{Department of Physics, Lehigh University, 16 Memorial Drive East, Bethlehem, PA 18015, USA}
\emailAdd{ara425@lehigh.edu}
\emailAdd{timm.wrase@lehigh.edu}

\abstract{We construct explicit periodic second-order localization corrections to scale-separated AdS$_4$ vacua of massive type IIA supergravity on $T^6/(\Z_2\times\Z_2)$. At first order, the corrections from different O6-plane families simply add. The interactions between intersecting families first enter at second order and could obstruct the continuation of the solution. We find explicit solutions for the warp factor, dilaton, full internal metric and required form fields, including the terms generated by these interactions. The metric at second order contains off-diagonal components. With a common regularization of the singular source terms, we verify the full field equations and the conditions required for a solution on a compact space. Using the pure-spinor formalism, we find that the second-order fields satisfy the supersymmetry conditions accessible at the retained orders.}

\begin{document}
\maketitle
\raggedbottom

\section{Introduction}
\label{sec:intro}
Scale separation is necessary for a string compactification to admit a lower-dimensional effective description over a substantial range of energies. In an AdS vacuum, the Kaluza--Klein scale must be well above the inverse AdS radius, while the string coupling and curvature must also remain under control. Whether these requirements can be satisfied simultaneously with stabilized moduli remains an important question in string theory; see~\cite{Coudarchet:2023mfs} for a review. Constraints from the ten-dimensional equations and the AdS distance conjecture sharpen this question~\cite{Gautason:2015tig,Lust:2019zwm}.

Massive type IIA orientifolds provide a concrete setting in which to address this question. Their flux-induced superpotentials allow supersymmetric AdS vacua with the bulk moduli stabilized at tree level~\cite{Grimm:2004ua,Derendinger:2004jn,Villadoro:2005cu}. The constructions of DeWolfe, Giryavets, Kachru and Taylor (DGKT) and related toroidal models further admit a large-flux regime with large volume and weak coupling~\cite{DeWolfe:2005uu, Camara:2005dc, Ihl:2006pp}. In the DGKT construction, the four-form fluxes can be increased without changing the O6 tadpole condition. This gives a parametric limit in which the smeared solution exhibits a hierarchy between the AdS and Kaluza--Klein scales. Understanding whether this regime survives the localization of the orientifold sources is essential to its ten-dimensional interpretation~\cite{Acharya:2006ne, Banks:2006hg}. The role of leading orientifold contributions in producing scale separation has been analyzed more generally in~\cite{Tringas:2025uyg}. The holographic interpretation of these vacua has also been investigated through their scalar spectra~\cite{Conlon:2021cjk,Apers:2022zjx} and by reconstructing brane descriptions through flux backtracking~\cite{Apers:2025pon}. Assuming that the dual CFT arises from a decoupled brane worldvolume theory, Bedroya and Steinhardt argue for a holographic obstruction to DGKT scale separation~\cite{Bedroya:2025ltj,Bedroya:2025fie}. Explicit near-horizon and low-energy absorption analyses instead find evidence for brane decoupling~\cite{Apers:2026lgi,Apers:2026ahu}, so this question remains under discussion.

The smeared approximation replaces the localized source densities by smooth distributions. This simplifies the Bianchi identities and field equations, but removes the spatial dependence induced by the sources. Restoring it generates warping, a varying dilaton and corrections to the internal metric and fluxes. Comparisons between smeared and localized solutions show that the accuracy of the approximation depends on the background and source configuration~\cite{Blaback:2010sj,Baines:2020dmu}. In particular, small corrections in the bulk do not by themselves settle the behavior near an orientifold locus; localized massive O6 solutions illustrate why the source region requires separate attention~\cite{Saracco:2012wc}.

For the scale-separated massive IIA vacua, a systematic treatment becomes possible at large flux. Junghans developed a systematic expansion around the smeared solution and obtained the leading localization corrections, illustrating the construction explicitly on the original DGKT $T^6/(\mathbb Z_3\times\mathbb Z_3)$ orientifold~\cite{Junghans:2020acz}. Marchesano, Palti, Quirant and Tomasiello constructed the leading supersymmetric localization corrections in the language of generalized geometry, including an explicit $T^6/(\Z_2\times\Z_2)$ example~\cite{Marchesano:2020qvg}. These results identify the first spatially dependent fields and provide the starting point for a nonlinear consistency test. Related perturbative localization methods have also been applied to scale-separated massive IIA AdS$_3$ vacua~\cite{Emelin:2022cac}, and to massless IIA backgrounds and their M-theory lifts~\cite{Cribiori:2021djm,VanHemelryck:2024bas}.

At first order the equations are linear, so the corrections from different O6 families simply add. At second order the families interact, and products of singular first-order fields enter the equations. There is also a global constraint: on a compact space, the source of a Poisson equation must have zero average. This imposes a solvability condition that must be checked separately from solving the nonzero Fourier modes. Emelin formulated the resulting second-order integral constraints and studied the leading near-source warp-factor and dilaton corrections, showing that these constraints can fix integration constants left undetermined at first order~\cite{Emelin:2024vug}. His analysis does not construct the full periodic second-order fields.

The source intersections make this test particularly relevant for the toroidal orientifold considered here. Emelin~\cite{Emelin:2024vug} pointed out that a cancellation in the second-order equations does not hold when two O6-plane families share a transverse direction. In the original DGKT model, however, the apparent O6-plane self-intersections disappear when a consistent orientifold involution is used and the orbifold singularities are resolved~\cite{Junghans:2023yue}. In the factorized $T^6/(\mathbb{Z}_2\times\mathbb{Z}_2)$ orientifold studied here, every pair of O6-plane families shares a transverse direction. We therefore evaluate the mixed second-order contributions explicitly. The metric ansatz is equally important: Bardzell, Federico, Smith and Wrase found an obstruction to fully localized, perpendicularly intersecting sources with a diagonal metric under the assumptions of their flat-space supersymmetric construction~\cite{Bardzell:2024anh}. This motivates retaining the off-diagonal tensor equations, while the different compact AdS setting requires an independent calculation.

In this paper we study the second-order continuation in the isotropic untwisted sector of the toroidal orientifold. We construct explicit periodic fields, including the off-diagonal metric components generated by the intersections, evaluate the full internal Einstein equation and its compact zero modes, and examine the coordinate freedom of the metric. We define the singular limit through heat regularization and its equivalent theta-function integral~\cite{Shandera:2003gx,Andriot:2019hay}. Our goal here is a consistency test of the two-derivative localization scheme. We do not attempt a calculation of corrected string-theory observables, for which higher-derivative effects must also be assessed~\cite{Junghans:2020acz,Andriot:2023fss}. We also test the second-order fields against the supersymmetry equations of~\cite{Marchesano:2020qvg}. The available conditions are satisfied; the next-order Romans-mass equation additionally fixes one combination of the second-order scalar constants. A complete supersymmetric continuation and global charge quantization require additional higher-order and harmonic data.

Section~\ref{sec:setup} fixes the geometry and conventions. Section~\ref{sec:expansion} specifies the expansion and regularized first-order fields. Section~\ref{sec:solutions} gives the explicit second-order solution. Section~\ref{sec:consistency} gives the compact solvability conditions and the domain of validity. Section~\ref{sec:supersymmetry} presents the supersymmetry tests and their remaining limitations. We conclude in section~\ref{sec:conclusions}. The appendices give the coefficient equations and verification details.

\section{Geometry and supergravity equations}
\label{sec:setup}
\subsection{Sources and conventions}

We use complex coordinates $z^a=x^a+i y^a$, with all real coordinates of period one. The orbifold generators, $\theta$ and $\omega$, and the spatial part of the orientifold involution, $\sigma$, act as
\begin{align}
\theta(z^1,z^2,z^3)&=(z^1,-z^2,-z^3)\,,&\omega(z^1,z^2,z^3)&=(-z^1,z^2,-z^3)\,,\\
\sigma(z^1,z^2,z^3)&=(\bar z^1,\bar z^2,\bar z^3)\,.
\end{align}
We order the real coordinates as $(x^1,y^1,x^2,y^2,x^3,y^3)$ and denote them collectively by $u^m$. The four O6 families have transverse coordinate triples
\begin{align}
U_0&=(y^1,y^2,y^3)\,,&U_1&=(y^1,x^2,x^3)\,,\label{eq:U0U1}\\
U_2&=(x^1,y^2,x^3)\,,&U_3&=(x^1,x^2,y^3)\,.
\end{align}
We choose the all-negative O6 model used in section 6.2 of~\cite{Marchesano:2020qvg}, with no additional D6-branes. Each family has eight components on the covering torus, at transverse coordinates zero or one half. We write $\delta_I$ for the sum of these eight coordinate delta functions. The bracket superscripts denote the expansion coefficients defined in section~\ref{sec:expansion}. The isotropic reference metric and basis forms are
\begin{equation}
g^{[0]}_{mn}=R^2\delta_{mn}\,,\qquad \omega_a=\dd x^a\wedge\dd y^a\,,\qquad \alpha_I=\dd U_I^1\wedge\dd U_I^2\wedge\dd U_I^3\,.
\end{equation}
We denote the inverse reference metric by $g_{[0]}^{mn}=R^{-2}\delta^{mn}$. The order of the one-forms fixes the signs of $\alpha_I$. Our $\alpha_I$ are the forms denoted $\beta^I$ in section 6.2 of~\cite{Marchesano:2020qvg}; their $\alpha_I$ instead denote the dual three-forms. The charge orientations in this basis are
\begin{equation}
(q_0,q_1,q_2,q_3)=(+1,-1,-1,-1)\,.
\end{equation}
These relative signs do not denote different O6 types. The positive density entering the negative-tension source term and the oriented charge current at leading order are respectively
\begin{equation}
\rho_I^{[0]}=\frac{\delta_I}{R^3}\,,\qquad \rho_{I,3}=q_I\delta_I\alpha_I\,.
\end{equation}
These objects have different metric dependence. In particular, the coordinate charge current does not acquire the factor $R^{-3}$ that appears in the scalar density.

For clarity, the oriented tangent volume forms divided by $R^3$ and the charge orientations are
\begin{center}
\begin{tabular}{ccl}
Family & $q_I$ & Oriented tangent volume form divided by $R^3$ \\
\hline
$0$ & $+1$ & $-\dd x^1\wedge\dd x^2\wedge\dd x^3$ \\
$1$ & $-1$ & $+\dd x^1\wedge\dd y^2\wedge\dd y^3$ \\
$2$ & $-1$ & $+\dd y^1\wedge\dd x^2\wedge\dd y^3$ \\
$3$ & $-1$ & $+\dd y^1\wedge\dd y^2\wedge\dd x^3$
\end{tabular}
\end{center}
We orient the six-torus by $\dd x^1\wedge\dd y^1\wedge\dd x^2\wedge\dd y^2\wedge\dd x^3\wedge\dd y^3$. Each tangent form in the table, wedged with $q_I\alpha_I$, gives this positive coordinate volume form. Thus the charge current has the Poincar\'e-dual orientation appropriate to that tangent orientation. All four tangent planes have the same calibration, $-\operatorname{Re}\prod_a(\dd x^a+i\dd y^a)$, with the factor $R^3$ restored for the metric volume.

We define the background geometric forms
\begin{equation}
J_{\mathrm{CY}}=R^2\sum_a\omega_a\,,\qquad \Omega_{\mathrm{CY}}=iR^3\prod_a(\dd x^a+i\dd y^a)\,.
\end{equation}
Then $\operatorname{Re}\Omega_{\mathrm{CY}}=R^3(\alpha_0-\alpha_1-\alpha_2-\alpha_3)$, and the common tangent calibration is $-\operatorname{Im}\Omega_{\mathrm{CY}}$.

\subsection{Field equations}

We work in string frame and define $\tau=e^{-\phi}$. Our notation $w$ for the warp factor and $\Pi_{I,mn}$ for the source tangent projectors follows~\cite{Junghans:2020acz}. The metric is
\begin{equation}
\dd s_{10}^2=w^2(u)\hat g_{\mu\nu}\dd x^\mu\dd x^\nu+g_{mn}(u)\dd u^m\dd u^n\,,\qquad \hat R_4=-12\,.
\end{equation}
The RR fields below are internal; the spacetime-filling four-form is represented by the internal six-form. We retain the charge and tension normalization $\mu>0$ explicitly. The Bianchi identities are~\cite{Junghans:2020acz,Emelin:2024vug}
\begin{align}
\dd F_0&=0\,,&\dd H_3&=0\,,&\dd F_2&=F_0H_3-\mu\sum_I\rho_{I,3}\,,\label{eq:bianchi}\\
\dd F_4&=H_3\wedge F_2\,,&\dd F_6&=0\,.
\end{align}
The equations of motion for the form fields are
\begin{align}
0&=\dd(w^4\star_6 F_2)+H_3\wedge w^4\star_6 F_4\,,\\
0&=\dd(w^4\star_6 F_4)+H_3\wedge w^4\star_6 F_6\,,\qquad 0=\dd(w^4\star_6 F_6)\,,\\
0&=\dd(w^4\tau^2\star_6 H_3)-F_0w^4\star_6 F_2-w^4\star_6 F_4\wedge F_2-w^4\star_6 F_6\wedge F_4\,.
\label{eq:formexact}
\end{align}
We use the exterior-algebra convention $A_p\wedge\star_6 A_p=|A_p|^2\operatorname{vol}_6$, with the Hodge star computed from the full internal metric.\footnote{The relative minus signs in the NSNS line of~\eqref{eq:formexact} differ from the plus signs printed in equation (2.6) of~\cite{Junghans:2020acz} and equation (3.8) of~\cite{Emelin:2024vug}. Our sign convention and the conventions printed in those references are related by replacing the star on internal $p$-forms with $\star' A_p=(-1)^{p(6-p)}\star_6 A_p$. This reverses the Hodge sign on $H_3$ and leaves the even RR forms unchanged, so our NSNS equation becomes the printed plus-sign form after multiplication by $-1$. We use $\star_6$ as defined above throughout; the kinetic-action variation fixing our relative signs is given in appendix~\ref{app:einsteinchecks}.}

For internal forms, we define the normalized inner product and norms by
\begin{align*}
A_p\cdot B_p&=\frac{1}{p!}A_{a_1\ldots a_p}B^{a_1\ldots a_p}\,,\qquad |A_p|^2=A_p\cdot A_p\,,\\
|A_p|^2_{mn}&=\frac{1}{(p-1)!}A_{ma_2\ldots a_p}A_n{}^{a_2\ldots a_p}\qquad(p\ge1)\,.
\end{align*}
For $p=0$ the tensor norm $|A_0|^2_{mn}$ is zero.

We use scalar combinations and the internal Einstein equation equivalent to equations (2.9)--(2.11) of~\cite{Junghans:2020acz}, in the form displayed in equation (3.11) of~\cite{Emelin:2024vug}:
\begin{align}
0={}&\tau\nabla^2\tau+(\partial\tau)^2+4\frac{\tau}{w}\partial_m\tau\partial^m w-\sum_{p=0,2,4,6}\frac{p-5}{4}|F_p|^2-\frac12\tau^2|H_3|^2-\frac34\tau\mu\sum_I\rho_I\,,\label{eq:tauexact}\\
0={}&\frac{\tau^2}{w}\nabla^2w+3\frac{\tau^2}{w^2}(\partial w)^2+2\frac{\tau}{w}\partial_m\tau\partial^m w-\frac{\tau^2}{4w^2}\hat R_4-\frac14\sum_{p=0,2,4,6}|F_p|^2+\frac14\tau\mu\sum_I\rho_I\,,\label{eq:wexact}\\
0={}&\tau^2R_{mn}+2\partial_m\tau\partial_n\tau-2\tau\nabla_m\partial_n\tau-4\frac{\tau^2}{w}\nabla_m\partial_nw-\frac12\tau^2|H_3|^2_{mn}\nonumber\\
&-\frac12\sum_{p=0,2,4,6}\left(|F_p|^2_{mn}-\frac12g_{mn}|F_p|^2\right)-\frac12\tau\mu\sum_I\left(\Pi_{I,mn}-\frac12g_{mn}\right)\rho_I\,.
\label{eq:einsteinexact}
\end{align}
The tensor $\Pi_{I,mn}$ is the metric projected onto the tangent space of the corresponding source. 

\section{Expansion and regularized first-order fields}
\label{sec:expansion}
\subsection{Large-flux expansion}
We take $n>0$ to be the common coefficient of the three basis four-forms $\omega_1\wedge\omega_2$, $\omega_1\wedge\omega_3$ and $\omega_2\wedge\omega_3$ in the leading $F_4$ flux. Thus $F_4=nF_4^{[0]}+\ldots$, with $F_4^{[0]}=\omega_1\wedge\omega_2 + \omega_1\wedge\omega_3 + \omega_2\wedge\omega_3$. The coefficient $n$ refers to the fixed coordinate basis and the norm of the flux also depends on the metric. The allowed values of $n$ must respect flux quantization. The smeared leading-order solution fixes the overall powers of $n$. The equations of motion determine the relative orders of the corrections, following equations (3.1)--(3.7) of~\cite{Junghans:2020acz} and equations (3.12)--(3.16) of~\cite{Emelin:2024vug}. The metric and scalar corrections have relative steps of $n^{-1}$ and the form fields have relative steps of $n^{-1/2}$, leading to the expansion
\begin{equation}
\begin{aligned}
g_{mn}&=n^{1/2}\left(g^{[0]}_{mn}+\frac{g^{[1]}_{mn}}{n}+\frac{g^{[2]}_{mn}}{n^2}+\cdots\right)\,,\\
\tau&=n^{3/4}\left(\tau_0+\frac{\tau^{[1]}}{n}+\frac{\tau^{[2]}}{n^2}+\cdots\right)\,,\\
w&=n^{3/4}\left(w_0+\frac{w^{[1]}}{n}+\frac{w^{[2]}}{n^2}+\cdots\right)\,,\\
F_2&=n^{1/2}\left(F_2^{[0]}+\frac{F_2^{[1]}}{n^{1/2}}+\frac{F_2^{[2]}}{n}+\frac{F_2^{[3]}}{n^{3/2}}+\cdots\right)\,,\\
F_4&=n\left(F_4^{[0]}+\frac{F_4^{[1]}}{n^{1/2}}+\frac{F_4^{[2]}}{n}+\cdots\right)\,,\\
H_3&=H_3^{[0]}+\frac{H_3^{[1]}}{n^{1/2}}+\frac{H_3^{[2]}}{n}+\cdots\,,\\
F_0&=f_0 \in \mathbb{Z}\,,\\
F_6&=0\,.
\end{aligned}
\label{eq:expansion}
\end{equation}
Square-bracket superscripts label ordinary expansion coefficients with integer indices. In particular, $[0]$ denotes the leading coefficient after the overall power of $n$ has been removed. We absorb the Taylor factorials into the coefficients, including derived coefficients and homogeneous constants. Our order-$k$ coefficients therefore equal the corresponding Taylor derivatives divided by $k!$, whereas equation (3.12) of~\cite{Emelin:2024vug} keeps this factor explicit. The source quantities accordingly expand as
\begin{equation*}
\rho_I=n^{-3/4}\left(\rho_I^{[0]}+\frac{\rho_I^{[1]}}{n}+\cdots\right)\,,\qquad
\Pi_{I,mn}=n^{1/2}\left(\Pi_{I,mn}^{[0]}+\frac{\Pi_{I,mn}^{[1]}}{n}+\cdots\right)\,.
\end{equation*}
The complementary background projector with lowered indices is $g^{[0]}_{\perp I,mn} = g^{[0]}_{mn}-\Pi^{[0]}_{I,mn}$. The smeared coefficient $F_2^{[0]}$ vanishes on our branch, so the first localized correction $F_2^{[1]}$ has coordinate weight $n^{1/2}n^{-1/2}=n^0$.

We measure the Romans mass in units of its flux quantum, so $f_0\in\mathbb Z$. The Bianchi identity $\dd F_0=0$ makes it spatially constant, and we hold this quantized flux fixed as $n$ grows. Thus $F_0=f_0$ has no localization corrections~\cite{DeWolfe:2005uu,Junghans:2020acz}. The six-form equation fixes $w^4\star_6F_6=c$ to a constant, and the four-form equation then gives $cH_3=-\dd(w^4\star_6F_4)$. Since the D6 tadpole requires $[H_3]\ne0$, taking cohomology gives $c=0$. Thus $F_6=0$ at all orders within the two-derivative equations used here.

For $F_6=0$ and vanishing smeared $F_2$, the Bianchi identities and form equations~\eqref{eq:bianchi}--\eqref{eq:formexact} require $F_4^{[1]}$ and $H_3^{[1]}$ to be closed and co-closed with respect to $g^{[0]}$. They are therefore harmonic. We choose their harmonic parts to vanish. With this choice, $F_2^{[2]}$ also obeys source-free closed and co-closed equations, and we choose its harmonic part to vanish as well. Thus the representative studied here has
\begin{equation}
F_2^{[0]}=F_2^{[2]}=F_4^{[1]}=H_3^{[1]}=0\,.
\end{equation}
The first nonzero form corrections beyond $F_2^{[1]}$ are $F_2^{[3]}$, $F_4^{[2]}$ and $H_3^{[2]}$. These terms have integer powers of $n$ after including the overall form weights. Higher-order coefficients are left undetermined. Below we solve the equations of motion for $g^{[2]}_{mn}$, $\tau^{[2]}$ and $w^{[2]}$ (up to stated homogeneous data), together with the form coefficients $F_2^{[3]}$, $F_4^{[2]}$ and $H_3^{[2]}$.

The isotropic smeared background is
\begin{align}
F_0&=f_0\,,\qquad F_6=0\,,\qquad H_3^{[0]}=\frac{8\mu}{f_0}\sum_Iq_I\alpha_I\,,\label{eq:background-H}\\
F_4^{[0]}&=\omega_1\wedge\omega_2+\omega_1\wedge\omega_3+\omega_2\wedge\omega_3\,,\label{eq:background-F4}\\
R^4&=\frac{5}{3f_0}\,,\qquad \tau_0=\frac{f_0^2R^3}{20\mu}\,,\qquad w_0=\frac{f_0R^3}{4\mu}\,.
\label{eq:background}
\end{align}
We take $f_0>0$ and retain the source normalization $\mu>0$ explicitly. The average coordinate density is eight for each family. Thus $f_0H_3^{[0]}=8\mu\sum_Iq_I\alpha_I$, which cancels the integrated charge current in~\eqref{eq:bianchi}. This is the D6-charge tadpole cancellation condition obtained by integrating the $F_2$ Bianchi identity over compact three-cycles~\cite{DeWolfe:2005uu,Junghans:2020acz}. Smearing preserves the total charge, so the same integrated condition constrains both the localized and smeared descriptions.
The background also obeys
\begin{equation}
H_3^{[0]}=\frac{2f_0}{5\tau_0}\operatorname{Re}\Omega_{\mathrm{CY}}\,,\qquad F_4^{[0]}=\frac{3f_0}{10}J_{\mathrm{CY}}\wedge J_{\mathrm{CY}}\,.
\end{equation}
These relations fix the relative signs in comparison with the toroidal solution of~\cite{Marchesano:2020qvg}.

\subsection{Green functions and the theta representation}
All torus integrals use the covering coordinates, with $\int_{T^6}\dd^6u\,f(u)=\int_0^1\dd u^1\cdots\int_0^1\dd u^6\,f(u)$. The coordinate volume is one. For each source family, let $\beta_I$ be the zero-average periodic Green function with
\begin{equation}
\Delta_0\beta_I=\frac{8-\delta_I}{R^3}\,,\qquad \int_{T^6}\dd^6u\,\beta_I=0\,,\qquad \Delta_0=R^{-2}\sum_m\partial_m^2\,.
\label{eq:green}
\end{equation}
We define this function by a heat limit, using the proper-time regularization of torus Green functions in~\cite{Shandera:2003gx, Andriot:2019hay} (see in particular section 3.2, equations (3.5)--(3.6), of the latter). For heat time $t>0$, set
\begin{equation}
\beta_{I,t}(u)=\frac{1}{2\pi^2R}\sum_{\ell\in\Z^3\setminus\{0\}}\frac{e^{-16\pi^2t|\ell|^2/R^2}e^{4\pi i\ell\cdot U_I}}{|\ell|^2}\,,\qquad \beta_I=\lim_{t\to0^+}\beta_{I,t}\,.
\label{eq:betafourier}
\end{equation}
The profiles converge in the $L^2$ norm $\|f\|_{L^2}=(\int_{T^6}\dd^6u\,|f|^2)^{1/2}$ and converge smoothly on compact subsets away from the sources. The undamped coefficients are square summable, but are not absolutely summable in three dimensions. All differentiated Fourier sums are first evaluated at positive $t$. The even lattice in~\eqref{eq:betafourier} follows from summing the eight sources at transverse coordinates zero and one half.

To compare with the regularized Green functions used in~\cite{Shandera:2003gx, Andriot:2019hay, Junghans:2020acz}, define the Jacobi function by $\theta_3(z,q)=\sum_{k\in\Z}q^{k^2}e^{2ikz}$. Integrating each damped Fourier coefficient gives the equivalent expression
\begin{equation}
\beta_{I,t}(u)=\frac{1}{2\pi^2R}\int_{16\pi^2t/R^2}^{\infty}\left[\prod_{j=1}^{3}\theta_3\bigl(2\pi U_I^j,e^{-s}\bigr)-1\right]\dd s\,.
\label{eq:thetaGreen}
\end{equation}
At positive $t$ the interchange of sum and integral is justified by absolute convergence. The subtraction removes the constant Fourier mode. Poisson resummation gives the small-$s$ Gaussian representation, so~\eqref{eq:thetaGreen} has a finite limit away from the source loci. Near a smooth point of one source, it gives $\beta_I\sim1/(4\pi r_I)$, with $r_I$ the proper transverse distance in $g^{[0]}$.

The same regulator must act on the coordinate sources. With $\delta_{I,t}=e^{t\Delta_0}\delta_I$, the defining relation is preserved exactly:
\begin{equation}
\Delta_0\beta_{I,t}=\frac{8-\delta_{I,t}}{R^3}\,,\qquad \int_{T^6}\dd^6u\,\beta_{I,t}=0\,.
\label{eq:regulatedPoisson}
\end{equation}
The heat parameter defines the singular limit and is independent of the large-flux parameter $n$. It is not a source-core size supplied by string theory.

\subsection{First-order solution and harmonic choice}
The first-order warp factor and dilaton corrections $w^{[1]}$ and $\tau^{[1]}$ can contain additive constants $C_w^{[1]}$ and $C_\tau^{[1]}$ that are not fixed at first order. For our second-order solution and harmonic choice, the compact constraints set these constants and the additive diagonal metric constants $C_{g,mm}^{[1]}$ to zero, as we will show below in section~\ref{sec:consistency}. The leading background values $w_0$ and $\tau_0$ remain those in~\eqref{eq:background}. The leading profiles are~\cite{Junghans:2020acz, Marchesano:2020qvg}
\begin{align}
w^{[1]}&=\frac{\mu w_0}{4\tau_0}\sum_I\beta_I\,,\qquad \tau^{[1]}=-\frac{3\mu}{4}\sum_I\beta_I\,,\label{eq:first-scalars}\\
\frac{g^{[1]}_{mm}}{g^{[0]}_{mm}}&=\frac{\mu}{2\tau_0}\sum_I\varsigma_{I,m}\beta_I\,,\qquad g^{[1]}_{mn}=0\quad(m\ne n)\,,\label{eq:first-metric}\\
F_2^{[1]}&=\mu\sum_Iq_I\star_{\perp I}^{[0]}\dd\beta_I\,.
\label{eq:first}
\end{align}
Here $\varsigma_{I,m}=+1$ in tangent directions and $-1$ in transverse directions. The transverse Hodge star has orientation $\alpha_I$, so $\dd\star_{\perp I}^{[0]}\dd\beta_I=(8-\delta_I)\alpha_I$ gives the localized two-form Bianchi identity. The first-order metric is diagonal and traceless but need not be divergence-free.

We choose zero harmonic corrections for the displayed forms and retain additive second-order scalar constants and six diagonal metric constants $C_{g,mm}^{[2]}=\int_{T^6}\dd^6u\,g^{[2]}_{mm}$, defined as metric means, whereas the scalar constants are additive coefficients rather than means. These second-order constants are undetermined by the retained bosonic equations; their values must be determined from the next-order integrated equations together with the harmonic flux data. Similarly, the first-order constants are fixed by the second-order integrated equations in section~\ref{sec:consistency}. A harmonic term $\delta n\sum_{a<b}\omega_a\wedge\omega_b$ in $F_4^{[2]}$, with $\delta n=O(1)$, can instead be absorbed into $n\mapsto n+\delta n$, accompanied by the corresponding re-expansion of the physical background factors $n^{1/2}g^{[0]}_{mn}$, $n^{3/4}\tau_0$ and $n^{3/4}w_0$ and of the higher-order terms. Thus changing the harmonic representative, for example when imposing Page periods, also changes the background-parameter expansion.

\section{Explicit second-order solution}
\label{sec:solutions}
We first determine $F_4^{[2]}$ and $H_3^{[2]}$, whose contractions with the background fluxes enter the second-order scalar and metric equations. We then solve for $w^{[2]},\tau^{[2]}$ and $g^{[2]}$, and determine $F_2^{[3]}$ from its Bianchi identity and equation of motion. Thus second order refers to the geometric expansion in~\eqref{eq:expansion}, with the field-dependent flux orders required by the equations. The isotropic choice concerns the reference radii and four-form fluxes; the localized metric corrections retain their full position and tensor dependence. We work with the bulk fields inherited from the covering torus and do not include exceptional-cycle or twisted-sector data. The pairwise O6 intersections lie on the fixed tori of $\theta$, $\omega$ or $\theta\omega$, hence at orbifold singularities of the quotient. Their fate after resolving the orbifold requires a separate analysis, an issue also raised for this model in~\cite{Cribiori:2026caf}. In related AdS$_3$ compactifications on $G_2$ orientifolds, resolving the orbifold removes the apparent orientifold intersections~\cite{Aggarwal:2026jan}.

Including all overall powers of $n$ in~\eqref{eq:expansion}, we solve the scalar equations and the two-form Bianchi identity through order $n^{-1}$, the internal Einstein equation through $n^{-1/2}$, the RR form equations through $n^{5/2}$, and the NSNS form equation through $n^{7/2}$. These powers refer to terms in the full equations, rather than to the bracket labels of individual field coefficients. Subsequent terms on this branch are suppressed by an additional $n^{-1}$. In the four-form Bianchi identity, $n\,\dd F_4^{[0]}$ vanishes because the background flux is closed, and we impose the first nontrivial coefficient at order $n^0$. The NSNS Bianchi identity holds through $n^{-1}$, the mass parameter $F_0$ is constant and the retained six-form vanishes.

The second-order sources contain products that cannot be interpreted separately at an O6 locus. In particular, $\beta_I^2\sim r_I^{-2}$ is locally integrable, whereas $(\partial\beta_I)^2\sim r_I^{-4}$ is not. At positive heat time the complete same-family combination obeys
\begin{equation}
(\partial\beta_{I,t})^2+\beta_{I,t}\Delta_0\beta_{I,t}=\frac12\Delta_0(\beta_{I,t}^2)\ \longrightarrow\ \frac12\Delta_0(\beta_I^2)\,.
\label{eq:groupedlimit}
\end{equation}
The limit defines the grouped distribution; it does not assign separate values to the two singular terms on the left. This is the point at which the near-source cancellation needed for the second-order integral constraints enters~\cite{Emelin:2024vug}.

We smooth the Green functions and the O6 densities at the same positive heat time, using~\eqref{eq:betafourier} and~\eqref{eq:regulatedPoisson}, and combine the quadratic terms before taking $t\to0^+$. The heat-smoothed profiles converge to $\beta_I$ in $L^2$. Writing the difference of the products as a sum of two terms and applying the Cauchy--Schwarz inequality to each gives
\begin{equation}
\|\beta_{I,t}\beta_{J,t}-\beta_I\beta_J\|_{L^1}\leq\|\beta_{I,t}-\beta_I\|_{L^2}\|\beta_{J,t}\|_{L^2}+\|\beta_I\|_{L^2}\|\beta_{J,t}-\beta_J\|_{L^2}\longrightarrow0\,.
\label{eq:productlimit}
\end{equation}
Here $\|f\|_{L^1}=\int_{T^6}\dd^6u\,|f|$. Thus the integral of the absolute difference tends to zero for both the squares $\beta_I^2$ and the mixed products $\beta_I\beta_J$, including at intersections. Their derivatives converge as periodic distributions. The Fourier kernels used below grow at most polynomially, so applying them to the products also gives well-defined distributional limits. This construction uses the complete combinations of terms, such as~\eqref{eq:groupedlimit}, in the two-derivative equations.

\subsection{Form fields}
We denote the background Hodge star by $\star_0$. Its first variation is
\begin{equation}
\star_1[g^{[1]}]A_p=\star_0\left[\frac12g_{[0]}^{mn}g^{[1]}_{mn}A_p-g_{[0]}^{ma}g^{[1]}_{an}\dd u^n\wedge\iota_{\partial_m}A_p\right]\,.
\label{eq:hodgevariation}
\end{equation}
The first term is the volume variation and the second accounts for the inverse metrics. The first-order metric is traceless, and its opposite deformations within each two-torus also give $\star_1[g^{[1]}]F_4^{[0]}=0$. We use this simplification in the four-form equation and solution below. The four-form and NSNS coefficients obey
\begin{align}
\dd F_4^{[2]}&=H_3^{[0]}\wedge F_2^{[1]}\,,\qquad \dd H_3^{[2]}=0\,,\\
\dd\star_0F_4^{[2]}&=-\dd\left(\frac{\mu}{\tau_0}\sum_I\beta_I\star_0F_4^{[0]}\right)\,,\label{eq:F4second}\\
\dd\star_0H_3^{[2]}&=-\dd\left(-\frac{\mu}{2\tau_0}\sum_I\beta_I\star_0H_3^{[0]}+\star_1[g^{[1]}]H_3^{[0]}\right)\nonumber\\
&\quad+\frac{1}{\tau_0^2}\left(f_0\star_0F_2^{[1]}+\star_0F_4^{[0]}\wedge F_2^{[1]}\right)\,.
\label{eq:Hsecond}
\end{align}
Every pair of transverse three-forms shares a one-form, so $\alpha_I\wedge\alpha_J=0$; also $J_{\mathrm{CY}}\wedge\alpha_I=0$. These identities establish source closure. The sources have zero mean because they are linear in derivatives of the Green functions. Thus their nonzero modes can be inverted without a harmonic obstruction.

We solve the form equations in Fourier space, where derivatives act by multiplication by momenta. For a periodic component $f(u)$, our Fourier convention and inverse transform are
\begin{align*}
\widehat f(N)&=\int_{T^6}\dd^6u\, f(u)e^{-2\pi iN\cdot u}\,,\qquad N\in\Z^6\,,\\
f(u)&=\sum_{N\in\Z^6}\widehat f(N)e^{2\pi iN\cdot u}\,.
\end{align*}
The six integers $N_m$ label a Fourier mode. For a differential form we transform each coordinate component and leave the basis one-forms fixed. In particular, $\widehat F_4^{[2]}(N)$ and $\widehat H_3^{[2]}(N)$ below are the Fourier coefficients of the corrections $F_4^{[2]}(u)$ and $H_3^{[2]}(u)$, reconstructed by the same inverse transform.

We write the momentum one-form as $\kappa_N=2\pi N_m\dd u^m$, so exterior differentiation on a Fourier mode acts as $i\kappa_N\wedge$. Throughout the component formulas, $|\kappa_N|^2=\sum_m(2\pi N_m)^2$ is the Euclidean coordinate norm and the norm computed with $g^{[0]}$ is $|\kappa_N|^2/R^2$. The operator $\iota_{\partial_m}$ contracts an index of a form, $(\iota_{\partial_m}A_p)_{a_2\ldots a_p}=A_{ma_2\ldots a_p}$. Thus contracting with the momentum vector uses $R^{-2}\sum_m\kappa_{N,m}\iota_{\partial_m}$, with the inverse-metric factor $g_{[0]}^{mn}=R^{-2}\delta^{mn}$ displayed explicitly.

To solve a Poisson equation with a zero-average periodic source $f$, we use $\Delta_0^{-1}f$, the unique zero-average solution whose Laplacian equals $f$. Each nonzero Fourier coefficient is divided by the Laplacian eigenvalue $-4\pi^2|N|^2/R^2$. The constant Fourier mode of the source must vanish separately: integrating the Poisson equation over the torus gives zero on the left-hand side by periodicity, and hence requires the source to have zero average. Explicitly,
\begin{equation}
\Delta_0^{-1}f=-\frac{R^2}{4\pi^2}\sum_{N\ne0}\frac{\widehat f(N)}{|N|^2}\exp(2\pi iN\cdot u)\,,\qquad \int_{T^6}\dd^6u\, f=0\,.
\label{eq:inverse}
\end{equation}
Applying this inversion to the Bianchi identities and the form equations~\eqref{eq:F4second} and~\eqref{eq:Hsecond} gives the following coefficients for $N\ne0$. Their zero modes vanish by our choice of harmonic corrections:
\begin{align}
\widehat F_4^{[2]}(N)=-\frac{iR^2}{4\pi^2|N|^2}\Bigg[&\frac{1}{R^2}\sum_m\kappa_{N,m}\iota_{\partial_m}\widehat{\left(H_3^{[0]}\wedge F_2^{[1]}\right)}(N)\nonumber\\
&+\kappa_N\wedge\star_0\widehat{\left\{-\dd\left(\frac{\mu}{\tau_0}\sum_I\beta_I\star_0F_4^{[0]}\right)\right\}}(N)\Bigg]\,.
\label{eq:F4solution}
\end{align}
\begin{align}
\widehat H_3^{[2]}(N)=-\frac{iR^2}{4\pi^2|N|^2}\kappa_N\wedge\star_0\Bigg[&\widehat{\left\{-\dd\left(-\frac{\mu}{2\tau_0}\sum_I\beta_I\star_0H_3^{[0]}+\star_1[g^{[1]}]H_3^{[0]}\right)\right\}}(N)\nonumber\\
&+\frac{1}{\tau_0^2}\widehat{\left(f_0\star_0F_2^{[1]}+\star_0F_4^{[0]}\wedge F_2^{[1]}\right)}(N)\Bigg]\,.
\label{eq:Hsolution}
\end{align}
Every source on the right is fixed by the first-order fields. At positive heat time these are smooth Fourier formulas. Taking $t\to0^+$ in the space of periodic distributions defines their limits. Contracting~\eqref{eq:F4solution} and~\eqref{eq:Hsolution} with the constant background forms gives the scalar sources below. The component and Hodge-star calculation, including the contribution of each term in the form equations, is given in appendix~\ref{app:formcontractions}:
\begin{equation}
\begin{aligned}
F_4^{[0]}\cdot F_4^{[2]}&=-\frac{12\mu f_0^2}{25\tau_0}\sum_I\beta_I\,,\\
H_3^{[0]}\cdot H_3^{[2]}&=\frac{16\mu f_0^2}{25\tau_0^3}\sum_I\beta_I\,.
\end{aligned}
\label{eq:signedcontractions}
\end{equation}
These contractions use the normalized form inner product of section~\ref{sec:setup} and their signs use the NSNS equation in~\eqref{eq:formexact}.

\subsection{Warp factor and dilaton}
Substituting the first-order profiles into the warp-factor equation~\eqref{eq:wexact} and the dilaton equation~\eqref{eq:tauexact}, and collecting the coefficients of $n^{-1}$, gives
\begin{align}
\Delta_0w^{[2]}={}&\frac{\mu^2w_0}{2\tau_0^2}\Delta_0\left[\frac18\sum_I\beta_I^2+\frac3{16}\left(\sum_I\beta_I\right)^2\right]\nonumber\\
&+\frac{\mu^2w_0}{8\tau_0^2}\partial_m\left(\sum_{I\ne J}\varsigma_{I,m}\beta_I\partial^m\beta_J\right)+\frac{27\mu w_0f_0^2}{50\tau_0^3}\sum_I\beta_I+\frac{w_0}{2\tau_0^2}F_4^{[0]}\cdot F_4^{[2]}\,,\label{eq:wsecond}\\
\Delta_0\tau^{[2]}={}&\frac{\mu^2}{2\tau_0}\Delta_0\left[\frac3{16}\left(\sum_I\beta_I\right)^2-\frac38\sum_I\beta_I^2\right]\nonumber\\
&-\frac{3\mu^2}{8\tau_0}\partial_m\left(\sum_{I\ne J}\varsigma_{I,m}\beta_I\partial^m\beta_J\right)-\frac{69\mu f_0^2}{50\tau_0^2}\sum_I\beta_I\nonumber\\
&-\frac{1}{2\tau_0}F_4^{[0]}\cdot F_4^{[2]}+\tau_0H_3^{[0]}\cdot H_3^{[2]}\,.
\label{eq:tausecond}
\end{align}
The form contractions in~\eqref{eq:signedcontractions} combine the terms linear in the $\beta_I$ into coefficients $3/10$ and $-1/2$ in the warp and dilaton equations. The quadratic sources are derivatives of the products defined above. Their averages vanish, as does the average of each linear Green-function source, so the inverse Laplacian applies. The periodic solutions are
\begin{align}
w^{[2]}={}&\frac{\mu^2w_0}{2\tau_0^2}\left[\frac18\sum_I\beta_I^2+\frac3{16}\left(\sum_I\beta_I\right)^2\right]+C_w^{[2]}\nonumber\\
&+\frac{\mu^2w_0}{8\tau_0^2}\Delta_0^{-1}\partial_m\left(\sum_{I\ne J}\varsigma_{I,m}\beta_I\partial^m\beta_J\right)+\frac{3\mu w_0f_0^2}{10\tau_0^3}\Delta_0^{-1}\sum_I\beta_I\,,\label{eq:wsolution}\\
\tau^{[2]}={}&\frac{\mu^2}{2\tau_0}\left[\frac3{16}\left(\sum_I\beta_I\right)^2-\frac38\sum_I\beta_I^2\right]+C_\tau^{[2]}\nonumber\\
&-\frac{3\mu^2}{8\tau_0}\Delta_0^{-1}\partial_m\left(\sum_{I\ne J}\varsigma_{I,m}\beta_I\partial^m\beta_J\right)-\frac{\mu f_0^2}{2\tau_0^2}\Delta_0^{-1}\sum_I\beta_I\,.
\label{eq:tausolution}
\end{align}
The constants $C_w^{[2]}$ and $C_\tau^{[2]}$ are additive homogeneous coefficients. They are not the averages of the full corrections, because the explicitly displayed quadratic terms have nonzero means. The linear inverse is itself an absolutely convergent series,
\begin{equation}
\Delta_0^{-1}\sum_I\beta_I=-\frac{R}{32\pi^4}\sum_I\sum_{\ell\in\Z^3\setminus\{0\}}\frac{e^{4\pi i\ell\cdot U_I}}{|\ell|^4}\,.
\end{equation}
For each pair $I<J$, let $c$ be its common transverse coordinate index. Then the mixed divergence can also be written entirely in terms of the product of two Green functions:
\begin{equation}
\partial_m\left(\varsigma_{I,m}\beta_I\partial^m\beta_J+\varsigma_{J,m}\beta_J\partial^m\beta_I\right) =\left(\Delta_0-\frac2{R^2}\partial_c^2\right)(\beta_I\beta_J)\,.
\label{eq:mixedscalarderivative}
\end{equation}
To see why this matters, consider the pair $(I,J)=(0,1)$ with common transverse coordinate $c=y^1$, see equation~\eqref{eq:U0U1}. Let $F_2^{[1],I}$ denote the $I$-th term in~\eqref{eq:first}. The two individual two-forms have no common index pair, so $F_2^{[1],0}\cdot F_2^{[1],1}=0$, whereas $\partial_m\beta_0\partial^m\beta_1=R^{-2}\partial_c\beta_0\partial_c\beta_1$ need not vanish. The same observation applies to every pair. Thus the flux-gradient identity used near intersections in~\cite{Emelin:2024vug} does not apply. Equation~\eqref{eq:mixedscalarderivative} displays the replacement: the shared direction changes the differential operator, while the complete mixed source remains a total derivative. It holds at positive heat time without imposing a source-free Poisson equation.

Near a smooth point of one source, the leading quadratic terms are
\begin{equation}
\frac{w^{[2]}}{w_0}\sim\frac{5\mu^2}{32\tau_0^2(4\pi r_I)^2}\,,\qquad \frac{\tau^{[2]}}{\tau_0}\sim-\frac{3\mu^2}{32\tau_0^2(4\pi r_I)^2}\,.
\label{eq:near-source-coefficients}
\end{equation}
The local Coulomb profile is $\mu\beta_I/\tau_0\sim\mu/(4\pi\tau_0r_I)$ in our normalization. The Coulomb coefficients $5/32$ and $-3/32$ follow by expanding $(1-\mu/(4\pi\tau_0nr_I))^{-1/4}$ and $(1-\mu/(4\pi\tau_0nr_I))^{3/4}$. With the factorial absorbed, they agree with the near-source results of~\cite{Emelin:2024vug}. Intersections have additional mixed singularities, so this isolated-source limit is not uniform as an intersection is approached.

To illustrate the interaction of families 0 and 1, use proper Cartesian coordinates for $g^{[0]}$. Let $c$ be their shared transverse coordinate, and let $\mathbf a,\mathbf b\in\mathbb R^2$ collect the other transverse coordinates of families 0 and 1, respectively. Their local distances and the Coulomb length are
\begin{align}
r_0&=\sqrt{c^2+|\mathbf a|^2}\,,\qquad r_1=\sqrt{c^2+|\mathbf b|^2}\,,\nonumber\\
\ell_{\mathrm C}&=\frac{\mu}{4\pi\tau_0n}\,,\qquad W=\frac{w}{n^{3/4}w_0}\,.
\label{eq:intersectioncoordinates}
\end{align}
Keep the common tangent coordinate away from the other two families. Inserting $\beta_0\simeq(4\pi r_0)^{-1}$ and $\beta_1\simeq(4\pi r_1)^{-1}$ into~\eqref{eq:wsolution}, and using~\eqref{eq:mixedscalarderivative}, gives the leading local terms
\begin{align}
W-1\simeq{}&\frac{\ell_{\mathrm C}}{4}\left(\frac1{r_0}+\frac1{r_1}\right)\nonumber\\
&+\ell_{\mathrm C}^2\left[\frac5{32}\left(\frac1{r_0^2}+\frac1{r_1^2}\right)+\frac5{16r_0r_1}-\frac14\partial_c^2\Delta_5^{-1}\frac1{r_0r_1}\right]+\cdots\,.
\label{eq:localintersectionwarp}
\end{align}
Here $\Delta_5=\partial_c^2+\sum_{j=1}^2(\partial_{a_j}^2+\partial_{b_j}^2)$ acts on all five transverse coordinates. The differentiated inverse is evaluated with the flat heat inverse before restricting to $c=0$. The mixed product coefficient is $3/16+1/8=5/16$: the first contribution comes from the local square in~\eqref{eq:wsolution}, and the second from the mixed divergence. The ellipsis contains less singular local terms and higher perturbative orders, including the linear-flux response, periodic images and the remaining source families.

\begin{figure}[t]
\centering
\includegraphics[width=\textwidth]{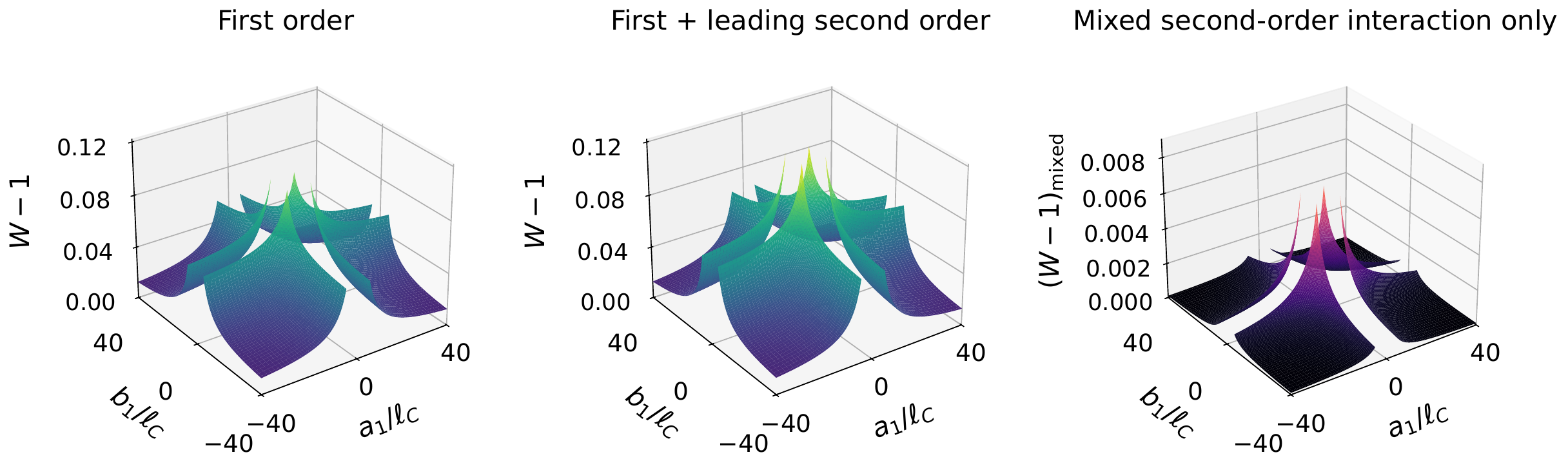}
\caption{Leading local warp-factor corrections near the intersection of O6 families 0 and 1, computed from~\eqref{eq:localintersectionwarp}. Coordinates are proper distances for the reference metric, in units of $\ell_{\mathrm C}=\mu/(4\pi\tau_0n)$, and the shared transverse coordinate is $c=0$. Left: the first-order contribution to $W-1$. Middle: first order plus the leading quadratic terms, on the same vertical scale. Right: $(W-1)_{\mathrm{mixed}}$, the mixed quadratic part of~\eqref{eq:localintersectionwarp}, on an expanded vertical scale. The strips $|a_1|<5\ell_{\mathrm C}$ or $|b_1|<5\ell_{\mathrm C}$, which contain the two source axes and their intersection, are omitted.}
\label{fig:O6intersection}
\end{figure}

Figure~\ref{fig:O6intersection} shows the slice $c=0$, $\mathbf a=(a_1,0)$ and $\mathbf b=(b_1,0)$. Both plotted distances are at least $5\ell_{\mathrm C}$. The displayed mixed term increases the quadratic correction and depends on both distances, as required for an intersection contribution.

Equation~\eqref{eq:localintersectionwarp} has the expected structure of a nonlinear continuation: the individual-source corrections are supplemented by terms depending on both source profiles. For comparable distances $r_0\sim r_1\sim r$, the displayed self and mixed terms are both of order $\ell_{\mathrm C}^2/r^2$, and remain smaller than the first-order terms when $\ell_{\mathrm C}/r\ll1$. Intersections therefore affect the second-order bulk fields even where the expansion is controlled.

\subsection{Internal metric}
\label{sec:explicitmetric}
To determine $g^{[2]}_{mn}$, we substitute~\eqref{eq:expansion} into the internal Einstein equation~\eqref{eq:einsteinexact} and collect the coefficient of $n^{-1/2}$, denoted by $\mathcal E^{[2]}_{mn}$ and displayed in~\eqref{eq:einsteinsecond}. This power counts the full equation, including the overall field factors; the corresponding term in the metric itself is $n^{-3/2}g^{[2]}_{mn}$. At this order the Ricci tensor contains terms linear in the unknown second-order correction $g^{[2]}_{mn}$ and terms quadratic in the known first-order correction $g^{[1]}_{mn}$. We denote precisely the terms linear in $g^{[2]}$ by $R^{\mathrm{lin}}_{mn}[g^{[2]}]$. The terms quadratic in $g^{[1]}$, together with the already determined scalar and form corrections, belong to the known source. The 21 independent metric equations that we have to solve therefore have the form
\begin{equation}
\mathcal E^{[2]}_{mn}=\tau_0^2R^{\mathrm{lin}}_{mn}[g^{[2]}]+\left.\mathcal E^{[2]}_{mn}\right|_{g^{[2]}=0}=0\,.
\label{eq:Einsteinlinearsplit}
\end{equation}

Diffeomorphism invariance means that different internal coordinates describe the same geometry. With the first-order fields fixed, a periodic second-order transformation generated by $\xi^m$ changes the metric coefficient by $\delta(g^{[2]}_{mn})=\partial_m\xi_n+\partial_n\xi_m$, where $\xi_n=g^{[0]}_{np}\xi^p$. We fix this freedom by imposing the de Donder gauge, the linear harmonic condition reviewed in section 6 of~\cite{Carroll:1997ar},
\begin{equation}
\partial^m g^{[2]}_{mn}-\frac12\partial_n\left(g_{[0]}^{mp}g^{[2]}_{mp}\right)=0\,.
\label{eq:gauge}
\end{equation}
The left-hand side changes by $\Delta_0\xi_n$, so the nonzero Fourier modes of $\xi_n$ can impose~\eqref{eq:gauge}. Its zero mode vanishes automatically on the torus. Periodic residual transformations satisfy $\Delta_0\xi_n=0$ and are constant, so they leave $g^{[2]}_{mn}$ unchanged. This fixes the coordinate freedom at second order while keeping the first-order fields in the coordinates chosen in~\eqref{eq:first-metric}.

The connection and Ricci variations~\eqref{eq:connectionlinear} and~\eqref{eq:Riccilinear} in appendix~\ref{app:geometric} apply to any linear perturbation of the flat background. Applying them to $g^{[2]}$ gives
\begin{align}
R^{\mathrm{lin}}_{mn}[g^{[2]}]={}&\frac{1}{2R^2}\Big[\sum_a\left(\partial_a\partial_mg^{[2]}_{an}+\partial_a\partial_ng^{[2]}_{am}-\partial_a^2g^{[2]}_{mn}\right)\nonumber\\
&-\partial_m\partial_n\sum_a g^{[2]}_{aa}\Big]\nonumber\\
={}&-\frac12\Delta_0g^{[2]}_{mn}+\frac12\partial_m\left[\partial^a g^{[2]}_{an}-\frac12\partial_n\left(g_{[0]}^{ab}g^{[2]}_{ab}\right)\right]\nonumber\\
&+\frac12\partial_n\left[\partial^a g^{[2]}_{am}-\frac12\partial_m\left(g_{[0]}^{ab}g^{[2]}_{ab}\right)\right]\,.
\label{eq:Ricciposition}
\end{align}
The last two terms are derivatives of the gauge condition~\eqref{eq:gauge} and vanish in that gauge. The 21 component equations therefore become Poisson equations.

Replacing each derivative $\partial_m$ by $i\kappa_{N,m}$ then gives, for $N\ne0$,
\begin{align}
\widehat{R^{\mathrm{lin}}_{mn}[g^{[2]}]}(N)=\frac{1}{2R^2}\Big[&|\kappa_N|^2\widehat g^{[2]}_{mn}-\kappa_{N,m}\sum_a\kappa_{N,a}\widehat g^{[2]}_{an}\nonumber\\
&-\kappa_{N,n}\sum_a\kappa_{N,a}\widehat g^{[2]}_{am}+\kappa_{N,m}\kappa_{N,n}\sum_a\widehat g^{[2]}_{aa}\Big]\,.
\label{eq:Riccisymbol}
\end{align}
In the gauge~\eqref{eq:gauge}, the last three terms cancel, leaving the Fourier-space Poisson operator
\begin{equation}
\widehat{R^{\mathrm{lin}}_{mn}[g^{[2]}]}(N)=\frac{|\kappa_N|^2}{2R^2}\widehat g^{[2]}_{mn}(N)\,,\qquad N\ne0\,.
\label{eq:Riccisymbolgauge}
\end{equation}
Using~\eqref{eq:Riccisymbolgauge} in~\eqref{eq:Einsteinlinearsplit} and solving its nonzero Fourier modes gives
\begin{equation}
\widehat g^{[2]}_{mn}(N)=-\frac{2R^2}{\tau_0^2|\kappa_N|^2}\left.\widehat{\mathcal E}^{[2]}_{mn}(N)\right|_{g^{[2]}=0}\,,\qquad N\ne0\,.
\label{eq:metricinversion}
\end{equation}
To obtain the component solutions, we evaluate the source $\left.\mathcal E^{[2]}_{mn}\right|_{g^{[2]}=0}$ in~\eqref{eq:einsteinsecond} using the first-order metric and the already determined scalar and form corrections. In particular, Fourier transforming~\eqref{eq:selftensorsource} and applying~\eqref{eq:metricinversion} gives the self-source contribution below. The mixed-source and linear-flux contributions follow from the same procedure, as detailed in appendix~\ref{app:metricsolution}. The zero modes must be checked separately; they are discussed together with the homogeneous coefficients in section~\ref{sec:consistency}.

To display the component solutions, we use the background complex structure acting on covectors, $J_m{}^a=(J_{\mathrm{CY}})_{mb}g_{[0]}^{ba}$, with matrix $\left(\begin{smallmatrix}0&1\\-1&0\end{smallmatrix}\right)$ on each $(x^a,y^a)$ pair. All explicit component sums below are Euclidean coordinate contractions, and the powers of $R$ are displayed.

The nonconstant metric coefficient is the sum of self-source, linear-flux and mixed-source terms. Its zero mode is $\widehat g^{[2]}_{mn}(0)=C_{g,mm}^{[2]}\delta_{mn}$, with no sum over $m$. For $N\ne0$, the self-source term is
\begin{equation}
\left.\widehat g^{[2]}_{mn}(N)\right|_{\rm self}=\frac{\mu^2R^2}{2\tau_0^2}\sum_I\left[\left(\frac14+\frac12\varsigma_{I,m}\right)\delta_{mn}+\frac{2\kappa_{N,m}\kappa_{N,n}}{|\kappa_N|^2}\right]\widehat{\beta_I^2}(N)\,.
\label{eq:metricselfexplicit}
\end{equation}

The flux response linear in the Green functions is
\begin{align}
\left.\widehat g^{[2]}_{mn}(N)\right|_{\rm linear}=\frac{\mu f_0^2R^4}{2\tau_0^3}\sum_I\Bigg[&\frac{4+2\varsigma_{I,m}}{5|\kappa_N|^2}\delta_{mn}\nonumber\\
&-\frac4{|\kappa_N|^4}\sum_{a,b}J_m{}^aJ_n{}^b\kappa_{N,a}\kappa_{N,b}\Bigg]\widehat\beta_I(N)\,.
\label{eq:metriclinearexplicit}
\end{align}

For a pair $I<J$, let $d$ denote its common tangent coordinate index, while $c$ denotes the common transverse coordinate. Their coordinate directions are
\begin{equation}
\begin{array}{c|rrrrrr}
(I,J)&(0,1)&(0,2)&(0,3)&(1,2)&(1,3)&(2,3)\\\hline
c&y^1&y^2&y^3&x^3&x^2&x^1\\
d&x^1&x^2&x^3&y^3&y^2&y^1
\end{array}\,.
\end{equation}
The mixed contribution is
\begin{align}
\left.\widehat g^{[2]}_{mn}(N)\right|_{\rm mixed} =\frac{\mu^2R^2}{2\tau_0^2}\sum_{I<J}\Bigg[&\left(\frac12-\frac{\kappa_{N,c}^2}{|\kappa_N|^2}\right)\delta_{mn}+\delta_{md}\delta_{nd}-\delta_{mc}\delta_{nc}\nonumber\\
&+\frac{\kappa_{N,c}}{|\kappa_N|^2}\left(\delta_{mc}\kappa_{N,n}+\delta_{nc}\kappa_{N,m}\right)\nonumber\\
&-\frac{2\kappa_{N,c}^2\kappa_{N,m}\kappa_{N,n}}{|\kappa_N|^4}\nonumber\\
&-\frac{f_0^2}{32\mu^2|\kappa_N|^2}\sum_{a,b}H^{[0]}_{cma}H^{[0]}_{cnb}\kappa_{N,a}\kappa_{N,b}\Bigg]\widehat{\beta_I\beta_J}(N)\,.
\label{eq:metricmixedexplicit}
\end{align}
The background components $H^{[0]}_{cma}$ include the coefficient $8\mu/f_0$ and the charge orientations in section~\ref{sec:setup}. The product has no dependence on the common tangent coordinate. Its Fourier coefficients are
\begin{equation}
\widehat{\beta_I\beta_J}(N)=\sum_{M\in\Z^6}\widehat\beta_I(M)\widehat\beta_J(N-M)\,.
\label{eq:metricproductconvolution}
\end{equation}
At positive heat time each factor carries its own damping from~\eqref{eq:betafourier}. For a fixed output mode, the unregulated self convolution has a three-dimensional $|M|^{-4}$ tail; a mixed convolution leaves at most one free summation direction, again with fourth-power decay. Thus these individual coefficients converge absolutely. This does not assert absolute convergence of the outer Fourier series reconstructing the metric. The latter is defined by the common heat limit using~\eqref{eq:productlimit}.

The mixed term gives a useful test of a diagonal metric ansatz. Take an even-lattice mode
\begin{equation}
\kappa_N=(0,\kappa_{y^1},0,0,\kappa_{x^3},\kappa_{y^3})\,,\qquad \kappa_{y^1}\kappa_{x^3}\kappa_{y^3}\ne0\,.
\end{equation}
No single-family profile has this Fourier support. Pairs $(0,3)$, $(1,2)$ and $(2,3)$ also lack the required support, while the $(0,2)$ and $(1,3)$ kernels have zero $x^2y^2$ entry. For the remaining pair $(0,1)$, $c=y^1$ and $d=x^1$. All Kronecker and longitudinal terms in this entry of~\eqref{eq:metricmixedexplicit} vanish. The surviving background components are $H^{[0]}_{y^1x^2x^3}=-8\mu/f_0$ and $H^{[0]}_{y^1y^2y^3}=8\mu/f_0$. Their product gives the full coefficient
\begin{equation}
\widehat g^{[2]}_{x^2y^2}(N)=\frac{\mu^2R^2}{\tau_0^2}\frac{\kappa_{x^3}\kappa_{y^3}}{|\kappa_N|^2}\widehat{\beta_0\beta_1}(N)\ne0\,.
\label{eq:offdiagonalexample}
\end{equation}
Under the orientifold reflection $y^a\mapsto-y^a$, the $x^2y^2$ metric component must be odd; since $\beta_0\beta_1$ is even and the kernel $\kappa_{x^3}\kappa_{y^3}/|\kappa_N|^2$ is odd under $\kappa_y\mapsto-\kappa_y$, the displayed component has precisely the required parity.

The convolution is positive on this support, since every nonzero Green-function Fourier coefficient is positive. This mode also gives a short check of the mixed Einstein equation. The scalar Hessians in the $x^2y^2$ entry vanish because both corresponding total momenta are zero. The remaining known source reduces to the Fourier transform of $\tfrac12\mu^2\partial_{y^3}\beta_0\partial_{x^3}\beta_1$. Since $\beta_0$ is independent of $x^3$ and $\beta_1$ is independent of $y^3$, its Fourier coefficient depends only on the total momenta. The complete equation is therefore
\begin{equation}
\widehat{\mathcal E}^{[2]}_{x^2y^2}(N)=\frac{\tau_0^2|\kappa_N|^2}{2R^2}\widehat g^{[2]}_{x^2y^2}(N)-\frac{\mu^2}{2}\kappa_{x^3}\kappa_{y^3}\widehat{\beta_0\beta_1}(N)=0\,.
\label{eq:mixedEinsteinexample}
\end{equation}
The first term is the unreduced linearized Ricci contribution in~\eqref{eq:Riccisymbol}; inserting~\eqref{eq:offdiagonalexample} cancels the source. This illustrates directly how the intersection generates a metric component that the Einstein equation requires.

A second-order coordinate change leaving the first-order fields fixed changes the metric by $\delta(\widehat g^{[2]}_{mn})=i(\kappa_{N,m}\widehat\xi_n+\kappa_{N,n}\widehat\xi_m)$, with the vector index lowered by the background metric. Both momenta $\kappa_{N,x^2}$ and $\kappa_{N,y^2}$ vanish in this mode, so this entry is unchanged. It cannot be removed by such a gauge transformation. This differs from the longitudinal off-diagonal entries in the self-source term~\eqref{eq:metricselfexplicit}: those arise from a Hessian with vanishing linearized Ricci tensor and impose the chosen gauge. The mixed component above has a nonzero Ricci contribution that is required to cancel its source. This gives a concrete perturbative reason to allow off-diagonal components in coordinates adapted to the source families, in line with the limitations of the diagonal ansatz discussed in~\cite{Bardzell:2024anh}.

For the nonzero modes, we verify the full internal Einstein equation with the unreduced Ricci operator~\eqref{eq:Riccisymbol} and the regulated source density~\eqref{eq:regulatedPoisson}. The mixed component in~\eqref{eq:mixedEinsteinexample} illustrates the check. We also verify the de Donder condition~\eqref{eq:gauge} independently. Appendix~\ref{app:metricsolution} gives the component checks.

\subsection{The next two-form coefficient}
\label{sec:ancillary}
The NSNS correction $H_3^{[2]}$ sources $F_2^{[3]}$ through the two-form Bianchi identity. This coefficient completes the retained form system and will provide a direct supersymmetry test of the second-order geometry in section~\ref{sec:supersymmetry}. Its equations are
\begin{align}
\dd F_2^{[3]}={}&f_0H_3^{[2]}\,,\label{eq:F23bianchi}\\
\dd\star_0F_2^{[3]}={}&-\dd\left[\frac{4w^{[1]}}{w_0}\star_0F_2^{[1]}+\star_1[g^{[1]}]F_2^{[1]}\right]\nonumber\\
&-H_3^{[2]}\wedge\star_0F_4^{[0]}-H_3^{[0]}\wedge\left[\star_0F_4^{[2]}+\star_1[g^{[1]}]F_4^{[0]}\right]\,.
\label{eq:F23eom}
\end{align}
In addition to $\star_1[g^{[1]}]F_4^{[0]}=0$ used in the four-form solution, the displayed fields obey $H_3^{[2]}\wedge J_{\mathrm{CY}}=0$ and $H_3^{[0]}\wedge\star_0F_4^{[2]}=0$. The self-source term inside the exterior derivative is closed. Applying the Fourier inversion used in~\eqref{eq:F4solution} and~\eqref{eq:Hsolution} to~\eqref{eq:F23bianchi} and~\eqref{eq:F23eom} gives the solution with zero harmonic part:
\begin{align}
\widehat F^{[3]}_{2,mn}(N)={}&\frac{i\mu f_0^2R^3}{\tau_0^2|\kappa_N|^2}\sum_{I,a}\kappa_{N,a}\left[\frac25q_I(\alpha_I)_{amn}-\sum_Jq_J(\alpha_J)_{amn}\right]\widehat\beta_I(N)\nonumber\\
&-\frac{i\mu f_0R}{8\tau_0}\sum_{I<J}\frac{\kappa_{N,c}}{|\kappa_N|^2}\sum_a\left[\kappa_{N,m}H^{[0]}_{cna}-\kappa_{N,n}H^{[0]}_{cma}\right]\kappa_{N,a}\widehat{\beta_I\beta_J}(N)\,.
\label{eq:F23explicit}
\end{align}
Its first line is the linear response required by $\dd(F_2^{[3]})=f_0H_3^{[2]}$. The second line is a closed mixed-source contribution. Its imaginary coefficients are odd under $N\mapsto-N$, so the physical two-form is real. As for the metric, derivatives of products are taken after forming the common-regulator products.

\section{Compact solvability and domain of validity}
\label{sec:consistency}
The constant Fourier modes of the second-order scalar and internal Einstein equations must be imposed separately. At positive heat time, the scalar divergences and Hessians integrate to zero, and the terms linear in the Green functions have zero mean. In the tensor equations the same-family terms are total derivatives. For two different families a derivative product can contribute to the mean only if the two momenta are opposite and lie along the shared transverse direction. The coefficients of these surviving terms cancel in each tensor component, as shown in appendix~\ref{app:metricsolution}. The localized profiles thus contribute no net source to these integrated equations.

With zero harmonic form corrections, the remaining mean equations constrain the additive constants in the first-order warp factor, dilaton and diagonal metric. We vary the metric and scalar constants in the smeared equations while holding fixed the coefficients of the flux forms in the coordinate basis built from $\dd x^a$ and $\dd y^a$. The flux forms are therefore unchanged, but their metric-dependent norms and contractions vary. This gives eight independent constraints (appendix~\ref{app:homogeneous}), implementing the mechanism described in~\cite{Emelin:2024vug}. They give
\begin{equation}
C_w^{[1]}=C_\tau^{[1]}=C_{g,mm}^{[1]}=0\qquad (m=1,\ldots,6)\,.
\label{eq:homogeneouszero}
\end{equation}
Only diagonal constant metric components survive the projections: components joining different two-tori are odd under $\theta$ or $\omega$, and each $g_{x^a y^a}$ is odd under $\sigma$. At second order, $C_w^{[2]}$, $C_\tau^{[2]}$ and the six $C_{g,mm}^{[2]}$ remain undetermined by the bosonic equations solved here. In the next-order integrated equations they enter with the same rank-eight linear operator as in appendix~\ref{app:homogeneous}, now with sources from lower-order products and harmonic flux terms. They must therefore be determined as part of that continuation, rather than fixed independently to zero.

Away from the O6 loci, on fixed compact subsets of the reference torus, the first- and second-order geometric corrections are suppressed by $n^{-1}$ and $n^{-2}$ relative to the background. The leading bulk ratio of the AdS radius to the internal radius remains $n^{1/2}w_0/R$. The corrected Kaluza--Klein spectrum remains to be computed. Near an isolated source, local control requires $\mu/(4\pi\tau_0 n r_I)\ll 1$. At intersections, the profiles of all approaching families and their mixed terms must be small. The distributional solution leaves the O6 cores unresolved~\cite{Saracco:2012wc,Junghans:2020acz}. Higher-derivative terms can enter at the same order in $1/n$ as the second-order corrections, so they must be included before extracting corrected observables~\cite{Andriot:2023fss,Emelin:2024vug}. At third order, terms such as $\beta_I^3\sim r_I^{-3}$ are not locally integrable in three transverse dimensions, so the present heat-limit argument does not extend term by term to that order.

\section{Supersymmetry at second order}
\label{sec:supersymmetry}
We now test the supersymmetry conditions accessible to the second-order fields at the perturbative accuracy specified in section~\ref{sec:solutions}, and derive a next-order condition on one combination of the homogeneous scalar coefficients. We use the $SU(3)\times SU(3)$ pure-spinor formalism~\cite{Grana:2005sn} in the $\theta=0$ branch of~\cite{Marchesano:2020qvg}, whose $\psi\to0$ limit is the smeared DGKT solution. Here $\theta$ is the pure-spinor phase, unrelated to the orbifold generator. The integrability results for calibrated sources relate supersymmetry to the bosonic equations when the requisite form equations and Bianchi identities are imposed~\cite{Koerber:2007hd}; we check the retained bosonic equations directly. The pure-spinor equations impose supersymmetry on the geometry and fluxes; the Bianchi identities are the same as in the bosonic construction and are retained at the same field-dependent orders. In particular, second-order geometry is accompanied by $F_2^{[3]}$.

\subsection{Pure-spinor equations and order counting}
In the notation of~\cite{Marchesano:2020qvg}, $v$ is a complex one-form, $\psi$ measures the relative angle of the internal spinors, and $J_\psi$ is the real two-form in the even pure spinor. Their flux-independent conditions (2.20a)--(2.20b) give
\begin{equation}
\operatorname{Re}v=-\frac{w}{2}\dd\log\bigl(w^3\tau\cos\psi\bigr)\,,\qquad \dd\bigl(w^3\tau\cos\psi\,J_\psi\bigr)=0\,.
\label{eq:susy-fluxfree}
\end{equation}
Here $e^A=\mu_{\mathrm{AdS}}w$, $e^{-\phi}=\tau$ and $\mu_{\mathrm{AdS}}=(n^{3/4}w_0)^{-1}$; the AdS parameter $\mu_{\mathrm{AdS}}$ is distinct from our source normalization $\mu$. The physical fluxes include the transformation $H_3=\hat H+\dd b$ and $F=e^{b\wedge}\mathcal F$, where $\hat H$ and $\mathcal F$ denote the NSNS and RR forms before this transformation and $b=\tan\psi\operatorname{Im}\omega$.

Let $(J,\Omega)$ denote the metric-compatible $SU(3)$ forms of one internal spinor. The odd pure spinor uses the complex two-form $\omega_\psi=(\operatorname{Re}\omega+i\operatorname{Im}\omega/\cos\psi)/\sin\psi$ (equation (2.19) of~\cite{Marchesano:2020qvg}), where $\omega$ is the complex two-form of the associated $SU(2)$ structure, distinct from the orbifold generator and the basis forms $\omega_a$. Since $\psi=O(n^{-3/2})$, through relative order $n^{-2}$ we can replace $\cos\psi J_\psi$ by $J$ and $iv\wedge\omega_\psi$ by $\Omega$ using their equation (2.12). At the orders used below their flux equations (2.22) reduce to
\begin{align}
H_3&=\frac{2}{w}\operatorname{Re}\Omega+\dd b\,,& f_0&=-J\cdot\dd(\tau\operatorname{Im}v)+5\frac{\tau}{w}\,,\nonumber\\
F_2&=-J\cdot\dd(\tau\operatorname{Im}\Omega)+\left(f_0-2\frac{\tau}{w}\right)b\,,\nonumber\\
F_4&=J\wedge J\left(\frac{f_0}{2}-\frac{\tau}{w}\right)+J\wedge\dd(\tau\operatorname{Im}v)+b\wedge\mathcal F_2\,.
\label{eq:susy-fluxes}
\end{align}
We use these expressions only through the indicated coefficient orders; terms suppressed by $\psi^2$ must be restored when extending them further. The contraction is $J\cdot=-\tfrac12J^{mn}\iota_{\partial_m}\iota_{\partial_n}$, with both indices raised using the metric. The leading transformation is $b=-\tfrac12\operatorname{Re}(\bar v\mathbin{\lrcorner}\Omega)$, with the contraction defined using that same metric. Although $b=O(n^{-1})$, both $\dd b$ and the $b$ term in $F_2$ contribute to the coefficients being tested.

The dictionary to the first-order solution in equations (6.2)--(6.10) of~\cite{Marchesano:2020qvg} is
\begin{equation}
g_s=\frac{1}{n^{3/4}\tau_0}\,,\qquad g_s\varphi=-\frac{\mu}{4\tau_0n}\sum_I\beta_I\,,\qquad \Delta_{\mathrm{CY}}=-n^{-1/2}\Delta_0\,.
\label{eq:susy-dictionary}
\end{equation}
Here $\varphi$ is their warp-factor potential and the last equality uses the physical leading metric $n^{1/2}g^{[0]}$. Write $J=n^{1/2}(J_{\mathrm{CY}}+n^{-2}J^{[2]}+\cdots)$ and $\Omega=n^{3/4}(\Omega_{\mathrm{CY}}+n^{-1}\Omega^{[1]}+n^{-2}\Omega^{[2]}+\cdots)$. We choose $J^{[1]}=0$, consistently with the fact that the first-order metric in~\eqref{eq:first-metric} is anti-hermitian with respect to the background complex structure. The first correction is explicitly
\begin{equation*}
\Omega^{[1]}=\frac{iR}{2}\sum_{a=1}^3g^{[1]}_{x^a x^a}\,\dd z^1\wedge\cdots\wedge\dd\bar z^a\wedge\cdots\wedge\dd z^3\,,
\end{equation*}
where only the $a$th factor is conjugated. This reproduces their toroidal deformation. Substitution in~\eqref{eq:susy-fluxes} recovers $F_2^{[1]}$, $F_4^{[2]}$ and $H_3^{[2]}$ in sections~\ref{sec:expansion} and~\ref{sec:solutions}; the NSNS agreement requires $\dd b$.

\subsection{Second-order geometry and flux conditions}
The $n^{-2}$ coefficient of $\log(w^3\tau)$ fixes the leading nonzero $v$. Using~\eqref{eq:wsolution} and~\eqref{eq:tausolution}, we find
\begin{align}
[\log(w^3\tau)]^{[2]}&=3\frac{w^{[2]}}{w_0}+\frac{\tau^{[2]}}{\tau_0}-\frac{3\mu^2}{8\tau_0^2}\left(\sum_I\beta_I\right)^2\nonumber\\
&=\frac{2\mu f_0^2}{5\tau_0^3}\Delta_0^{-1}\sum_I\beta_I+3\frac{C_w^{[2]}}{w_0}+\frac{C_\tau^{[2]}}{\tau_0}\,.
\label{eq:susy-log}
\end{align}
All quadratic and nonlocal mixed terms cancel in this combination. Thus~\eqref{eq:susy-fluxfree} fixes
\begin{equation}
\operatorname{Re}v=-\frac{\mu w_0f_0^2}{5\tau_0^3n^{5/4}}\dd\Delta_0^{-1}\sum_I\beta_I+O(n^{-9/4})\,.
\label{eq:susy-v}
\end{equation}
Its imaginary part is fixed by requiring $v$ to have type $(1,0)$. This is also the solution implied by the first-order Romans-mass equation and equation (6.10) of~\cite{Marchesano:2020qvg}. In particular, the leading $b$ coefficient is
\begin{equation*}
b^{[1]}=\frac{\mu w_0f_0^2}{5\tau_0^3}\,\iota_{(\dd\Delta_0^{-1}\sum_I\beta_I)^{\sharp_0}}\operatorname{Re}\Omega_{\mathrm{CY}}\,,\qquad b=n^{-1}b^{[1]}+\cdots\,,
\end{equation*}
where $\sharp_0$ raises the one-form index with $g_{[0]}^{mn}$.

The metric fixes the $(1,1)$ part of $J^{[2]}$ by algebraic compatibility. In matrix notation, with $P_{11}$ the background $(1,1)$ projection on antisymmetric two-tensors, the compatibility and closure conditions are
\begin{align}
J^{[2](1,1)}&=P_{11}\!\left(J_{\mathrm{CY}}g_{[0]}^{-1}g^{[2]}\right)-\frac12J_{\mathrm{CY}}\left(g_{[0]}^{-1}g^{[1]}\right)^2\,,\nonumber\\
\dd J^{[2]}&=-\frac{2\mu f_0^2}{5\tau_0^3}\dd\Delta_0^{-1}\sum_I\beta_I\wedge J_{\mathrm{CY}}\,.
\label{eq:susy-J}
\end{align}
In the first line the antisymmetric part is understood before projection. For each nonzero Fourier mode, the second line fixes the remaining $(2,0)+(0,2)$ part: the map $Z\mapsto\kappa_N\wedge Z$ is injective on these real two-forms. The closure equation has 20 three-form components for six real $(2,0)+(0,2)$ coefficients. Substitution of~\eqref{eq:metricselfexplicit}, \eqref{eq:metriclinearexplicit} and~\eqref{eq:metricmixedexplicit} satisfies this overdetermined system in the four linear, four self-source and six mixed sectors. Thus compatibility and closure are a nontrivial test of the computed metric, rather than an automatic consequence of choosing an almost complex structure. These checks are performed mode by mode, replacing $\dd$ by $i\kappa_N\wedge$ and solving for the six coefficients of $Z$ in the basis $\operatorname{Re}(\dd z^a\wedge\dd z^b)$, $\operatorname{Im}(\dd z^a\wedge\dd z^b)$, $a<b$. For example, in the linear sector, direct substitution gives, for $N\ne0$,
\begin{equation*}
\widehat{J^{[2](1,1)}}_{\rm lin}(N)
+\widehat{\bigl[\log(w^3\tau)\bigr]^{[2]}}_{\rm lin}(N)
J_{\mathrm{CY}}
=\frac{\mu f_0^2R^4}{\tau_0^3|\kappa_N|^4}
\sum_I\widehat\beta_I(N)\,\kappa_N\wedge(J\kappa_N)\,,
\end{equation*}
where $(J\kappa_N)_m=J_m{}^a\kappa_{N,a}$ uses the background complex structure defined in section~\ref{sec:explicitmetric}. Wedge multiplication by $\kappa_N$ vanishes, so $Z_{\rm lin}=0$.

The full solution and metric determine $\mathcal I^m{}_n=g^{mp}J_{pn}$. A concrete choice of $(1,0)$ coframe is $\eta^a_m=\tfrac12[(\dd z^a)_m+i(\dd z^a)_p\mathcal I^p{}_m]$. Expanding $\Omega=i n^{3/4}R^3\mathcal N\,\eta^1\wedge\eta^2\wedge\eta^3$ through relative order $n^{-2}$ determines $\Omega^{[2]}$, with the real factor $\mathcal N=1+O(n^{-2})$ fixed by $\tfrac{i}{8}\Omega\wedge\bar\Omega=\tfrac16J^3$. The flux comparison below fixes the remaining phase and admits zero phase correction. This constructs the nonzero modes of the structure; constant structure data must also satisfy the harmonic flux equations at the retained orders.

Expanding the two-form equation gives a direct comparison with the two-form coefficient computed in section~\ref{sec:ancillary}. Using the coefficient above gives
\begin{align}
F_2^{[3]}={}&-J_{\mathrm{CY}}\cdot\dd\left(\tau_0\operatorname{Im}\Omega^{[2]}+\tau^{[1]}\operatorname{Im}\Omega^{[1]}+\tau^{[2]}\operatorname{Im}\Omega_{\mathrm{CY}}\right)\nonumber\\
&+\frac{3f_0}{5}b^{[1]}\,.
\label{eq:susy-F2}
\end{align}
The first-order variation $(J\cdot)^{[1]}$ vanishes. Indeed, $J^{[1]}=0$ and the diagonal first-order metric is anti-hermitian: $g^{[1]}_{x^a x^a}+g^{[1]}_{y^a y^a}=0$ for each two-torus. The two inverse-metric variations therefore cancel in every nonzero component of $J_{\mathrm{CY}}^{mn}$. This expression agrees with~\eqref{eq:F23explicit} in all sectors. The constructed corrections are orbifold invariant and obey the orientifold projections $\sigma^*J=-J$, $\sigma^*\Omega=-\bar\Omega$ and $\sigma^*v=\bar v$.

\subsection{A next-order homogeneous condition and remaining tests}
The Romans-mass equation also relates different geometric orders. Since $J\cdot\sim n^{-1/2}$ and $\tau\sim n^{3/4}$, the coefficient of $v$ at $n^{-9/4}$ enters this equation at $n^{-2}$. Expanding the first equation in~\eqref{eq:susy-fluxfree} gives
\begin{align}
\operatorname{Re}v={}&-\frac{w_0}{2n^{5/4}}\dd[\log(w^3\tau)]^{[2]}\nonumber\\
&-\frac{w_0}{2n^{9/4}}\left(\dd[\log(w^3\tau\cos\psi)]^{[3]}+\frac{w^{[1]}}{w_0}\dd[\log(w^3\tau)]^{[2]}\right)+O(n^{-13/4})\,.
\label{eq:susy-v-next}
\end{align}
Here $[\log(w^3\tau\cos\psi)]^{[3]}=[\log(w^3\tau)]^{[3]}-\tfrac12[\psi^2]^{[3]}$. Although the first term involves a combination of third-order scalar coefficients, the leading spinor-angle term is already determined by~\eqref{eq:susy-v}:
\begin{equation}
[\psi^2]^{[3]}=\frac{w_0^2}{4R^2}\sum_m\left(\partial_m[\log(w^3\tau)]^{[2]}\right)^2\,.
\label{eq:susy-psi-leading}
\end{equation}
The nonzero modes of the Romans-mass equation determine $[\log(w^3\tau\cos\psi)]^{[3]}$ through a Poisson equation with sources fixed by the second-order fields. This supplies the required scalar combination directly from supersymmetry, without solving for the complete third-order geometry. Its zero mode gives a necessary condition for this higher-order continuation, rather than an additional equation at the retained bosonic orders. Because $J_{\mathrm{CY}}$ is constant and $(J\cdot)^{[1]}=0$, the differentiated term in the Romans-mass equation has zero mean at order $n^{-2}$. Substituting~\eqref{eq:wsolution} and~\eqref{eq:tausolution} into the remaining ratio gives
\begin{equation*}
0=\int_{T^6}\dd^6u\,\left[\frac{\tau}{w}\right]^{[2]}=\frac{\tau_0}{w_0}\left[\frac{C_\tau^{[2]}}{\tau_0}-\frac{C_w^{[2]}}{w_0}+\frac{\mu^2}{2\tau_0^2}\sum_{I<J}\int_{T^6}\dd^6u\,\beta_I\beta_J\right]\,.
\end{equation*}
The self-source means cancel. For $I\ne J$, only opposite Fourier modes along the shared transverse direction contribute, giving $\int_{T^6}\dd^6u\,\beta_I\beta_J=2\zeta(4)/(4\pi^4R^2)=1/(180R^2)$. The six mixed pairs therefore give
\begin{equation}
\frac{C_\tau^{[2]}}{\tau_0}-\frac{C_w^{[2]}}{w_0}=-\frac{\mu^2}{60R^2\tau_0^2}\,.
\label{eq:susy-constants}
\end{equation}
Thus a supersymmetric continuation restricts the bosonic family; setting both additive scalar constants to zero fails this condition. The other scalar combination remains undetermined at the retained order. Once~\eqref{eq:susy-constants} is imposed, the Romans-mass equation has a periodic solution for the scalar combination in~\eqref{eq:susy-v-next}. An independent third-order bosonic calculation could check this prediction, but is not needed to determine it.

The second-order metric and scalars therefore admit the required compatible structure, and the pure-spinor expressions reproduce the retained form coefficients. These are the supersymmetry tests accessible at the retained orders specified in section~\ref{sec:solutions}; equation~\eqref{eq:susy-constants} is a necessary next-order solvability condition. A complete supersymmetric continuation also requires higher-order geometry and compatible harmonic flux data. The next two-form Bianchi identity $\dd F_2^{[5]}=f_0H_3^{[4]}$ requires third-order geometry and remains open.

\section{Conclusions}
\label{sec:conclusions}
We have constructed explicit periodic second-order localization corrections for a toroidal massive type IIA orientifold with scale separation. The solution gives the warp factor, dilaton, full internal metric and the form coefficients required by their equations. It continues the first-order construction of~\cite{Junghans:2020acz,Marchesano:2020qvg} and supplies spatially resolved fields for the nonlinear consistency problem raised in~\cite{Emelin:2024vug}.

The source intersections are central to the calculation. Every pair shares a transverse direction, so the flux-gradient cancellation used for a different class of intersections does not apply. We have instead evaluated the mixed scalar and tensor terms and shown how their complete sources fit into the periodic solution. One intersection-sourced off-diagonal metric component survives second-order coordinate changes with fixed first-order data. This makes explicit why a diagonal ansatz would miss this branch, and gives a perturbative comparison with the restrictions on diagonal metrics found in~\cite{Bardzell:2024anh}.

A first-order solution need not extend to higher orders, since additional consistency conditions must be satisfied~\cite{Emelin:2024vug}. Here the mixed terms satisfy the tensor equations and the compact zero-mode constraints after the regulated sources are combined. Thus, within two-derivative supergravity and our common heat prescription, the intersections produce no obstruction at the retained order. Together with the required off-diagonal response, this provides a nontrivial test of continuing the localized solution beyond a linear superposition of individual sources.

To define the quadratic sources at the O6 loci, we smooth the first-order Green functions and the O6 source densities with the same heat time, as in~\eqref{eq:betafourier} and~\eqref{eq:regulatedPoisson}. The Green functions have the equivalent theta representation~\eqref{eq:thetaGreen} as in~\cite{Shandera:2003gx,Andriot:2019hay}. We combine the quadratic terms before taking the heat time to zero, so that the cancellations give well-defined distributions. In this limit we verify the full internal Einstein equation and independently check that the second-order metric satisfies the de Donder gauge fixing condition.

Integrating the second-order scalar and internal Einstein equations over the torus gives the zero-mode solvability conditions, following the approach of~\cite{Emelin:2024vug}. With our heat prescription and our choice of zero harmonic corrections to the form fields, these conditions set the first-order additive constants in the warp factor, dilaton and diagonal metric to zero.

The second-order fields pass the supersymmetry tests accessible at the retained orders, extending the first-order construction of~\cite{Marchesano:2020qvg}. The corrected metric admits compatible internal structure forms, and the pure-spinor equations reproduce the computed flux corrections. The next-order Romans-mass equation imposes a necessary condition on one combination of the second-order scalar constants and determines one higher-order scalar combination.

These results establish a continuation within two-derivative supergravity. They do not determine the source cores or a quantized global background. In particular, the physics directly at an intersection remains undetermined: the bulk expansion loses control near the source cores, the orbifold singularities require additional data, and higher-derivative terms can compete at the same order in $1/n$~\cite{Andriot:2023fss,Emelin:2024vug}. Higher-derivative effects and quantum corrections to brane dynamics require separate treatment before using these fields to infer corrected physical observables~\cite{Saracco:2012wc, Junghans:2020acz, Andriot:2023fss, Montero:2024qtz}. Revello and Van Hemelryck find~\cite{Revello:2026eqp, Revello:2026yla} that DGKT-CFI vacua on the $\Z_2\times\Z_2$ orbifold generically violate a proposed holographic constraint on cubic scalar couplings~\cite{Bobev:2025yxp}, whose implications are not settled by the supergravity consistency checks performed here. A natural next step is to solve the remaining higher-order supersymmetry conditions and match the harmonic data to quantized fluxes. Extending the calculation to unequal bulk fluxes, as in~\cite{Tringas:2023vzn}, and identifying observables insensitive to the source regions would further clarify the implications for scale-separated compactifications~\cite{Coudarchet:2023mfs}.

\subsection*{Acknowledgments}
We would like to thank Daniel Junghans and George Tringas for useful discussions and valuable feedback on an earlier version of this draft. This work is supported in part by the NSF grant PHY-2609419. This work was supported in part by Anthropic's AI for Science program. We acknowledge OpenAI for providing complimentary access to ChatGPT through the ChatGPT for Academic Researchers program.

\paragraph*{AI assistance.}
The authors used several large language models to assist with writing, editing, and checking calculations in this paper. The authors have carefully reviewed all such contributions and take full responsibility for the content, accuracy, and conclusions of the work.

\appendix
\section{Conventions and geometric variations}
\label{app:variation}
\subsection{NSNS sign convention}
\label{app:einsteinchecks}

The sign of the NSNS equation must be matched to the definition of the Hodge star. With $A_p\wedge\star_{10}A_p=|A_p|^2\operatorname{vol}_{10}$, the relevant kinetic terms are $-\frac12\tau^2H_3\wedge\star_{10}H_3-\frac12F_2\wedge\star_{10}F_2-\frac12F_4\wedge\star_{10}F_4$. We have $H_3=\dd B_2$, $F_2=\dd C_1+f_0B_2$ and $F_4=\dd C_3-H_3\wedge C_1+\frac12f_0B_2\wedge B_2$. In the internal-flux sector with $F_6=0$, variation with respect to $B_2$, followed by integration by parts and use of the $C_3$ equation, gives
\begin{equation}
\delta_{B_2}S=\int\delta B_2\wedge\left[\dd(\tau^2\star_{10}H_3)-f_0\star_{10}F_2-F_2\wedge\star_{10}F_4\right]\,.
\end{equation}
Indeed, $\delta F_4=\delta B_2\wedge F_2-\dd(\delta B_2\wedge C_1)$; the last term contributes only the $C_3$ equation. The topological term does not contribute in this purely internal sector. We suppress the common action normalization and boundary terms. This fixes the relative minus signs in~\eqref{eq:formexact}.

\subsection{Connection, curvature and source variations}
\label{app:geometric}

The raised metric perturbation is $g_{[1]}^{mn}=g_{[0]}^{ma}g_{[0]}^{nb}g^{[1]}_{ab}$, so the inverse metric has first variation $-g_{[1]}^{mn}$. For the flat background, the connection coefficients are
\begin{align}
\Gamma^{[1]p}{}_{mn}&=\frac12g_{[0]}^{pq}(\partial_mg^{[1]}_{nq}+\partial_ng^{[1]}_{mq}-\partial_qg^{[1]}_{mn})\,,\label{eq:connectionlinear}\\
\Gamma^{[2]p}{}_{mn}&=\frac12g_{[0]}^{pq}(\partial_mg^{[2]}_{nq}+\partial_ng^{[2]}_{mq}-\partial_qg^{[2]}_{mn})\nonumber\\
&\quad-\frac12g_{[1]}^{pq}(\partial_mg^{[1]}_{nq}+\partial_ng^{[1]}_{mq}-\partial_qg^{[1]}_{mn})\,.\label{eq:connectionsecond}
\end{align}
The corresponding Ricci coefficients, before imposing a gauge condition, are
\begin{align}
R^{[1]}_{mn}&=\partial_p\Gamma^{[1]p}{}_{mn}-\partial_n\Gamma^{[1]p}{}_{mp}\,,\label{eq:Riccilinear}\\
R^{[2]}_{mn}&=\partial_p\Gamma^{[2]p}{}_{mn}-\partial_n\Gamma^{[2]p}{}_{mp}+\Gamma^{[1]p}{}_{pq}\Gamma^{[1]q}{}_{mn}-\Gamma^{[1]p}{}_{nq}\Gamma^{[1]q}{}_{mp}\,.\label{eq:Riccisecond}
\end{align}
The Hodge-star variation is given in~\eqref{eq:hodgevariation}. For fixed source embeddings the scalar density varies as
\begin{equation}
\rho_I^{[1]}=-\frac12g_{[0],\perp I}^{ab}g^{[1]}_{ab,\perp I}\rho_I^{[0]}\,.
\label{eq:sourcedensityvariation}
\end{equation}
These variations must be included alongside the tangent-projector variation in the source stress tensor. Terms containing a singular metric coefficient and a source delta function use the common heat prescription in section~\ref{sec:solutions}.

After extracting the overall field weights, write $A_p=A_p^{[0]}+n^{-1/2}A_p^{[1]}+n^{-1}A_p^{[2]}+\cdots$. The ordinary coefficient of its squared norm is
\begin{equation}
[n^{-1}]|A_p|^2=2A_p^{[0]}\cdot A_p^{[2]}+|A_p^{[1]}|^2-g_{[1]}^{mn}|A_p^{[0]}|^2_{mn}\,.
\label{eq:norm}
\end{equation}
Here $[n^{-1}]$ means coefficient extraction. For $F_4$ and $H_3$ the chosen branch has $A_p^{[1]}=0$; for $F_2$ it has $A_p^{[0]}=0$, so this expression retains the squared norm of the leading localized two-form. In particular, the NSNS contribution to the dilaton equation has coefficient
\begin{align}
[n^{-1}]\left(-\tfrac12\tau^2|H_3|^2\right)={}&-\tau_0\tau^{[1]}|H_3^{[0]}|^2-\tau_0^2H_3^{[0]}\cdot H_3^{[2]}\nonumber\\
&+\frac{\tau_0^2}{2}g_{[1]}^{mn}|H_3^{[0]}|^2_{mn}\,.
\label{eq:Hcorrection}
\end{align}
The fields in this coefficient extraction have their overall powers of $n$ removed. Coefficient extraction returns the expansion coefficient defined in~\eqref{eq:expansion}. The scalar and tensor equations below are likewise written for the coefficients of their physical powers of $n$.

\subsection{Contractions of the form corrections}
\label{app:formcontractions}
We evaluate~\eqref{eq:signedcontractions} directly from the Fourier solutions. Consider a contribution proportional to $\widehat\beta_I(N)$, with $N\ne0$ supported in $U_I$. Since the background forms are constant, contraction commutes with the Fourier transform. With the coordinate-wedge components in~\eqref{eq:background-H}--\eqref{eq:background-F4}, the normalized inner product gives
\begin{align}
F_4^{[0]}\cdot\left(\kappa_N\wedge\sum_m\kappa_{N,m}\iota_{\partial_m}F_4^{[0]}\right)&=\frac{2|\kappa_N|^2}{R^8}\,,\nonumber\\
H_3^{[0]}\cdot\left(\kappa_N\wedge\sum_m\kappa_{N,m}\iota_{\partial_m}H_3^{[0]}\right)&=\frac{128\mu^2|\kappa_N|^2}{f_0^2R^6}\,.
\label{eq:backgroundmomentumcontractions}
\end{align}
Each coordinate occurs in two of the three background four-form components and in two of the four background three-form components. Distinct components have no common term after one contraction, which gives the diagonal momentum sums above. The Hodge identity
\begin{equation}
\star_0(\kappa_N\wedge\star_0 A_p)=R^{-2}\sum_m\kappa_{N,m}\iota_{\partial_m}A_p
\label{eq:Hodgecontractionidentity}
\end{equation}
then reduces the contractions of~\eqref{eq:F4solution} and~\eqref{eq:Hsolution} to these coordinate sums.

For the four-form, the equation-of-motion source $-\dd[(\mu/\tau_0)\sum_I\beta_I\star_0F_4^{[0]}]$ gives the first term below. The Bianchi source $H_3^{[0]}\wedge F_2^{[1]}$ gives the second:
\begin{align}
\widehat{F_4^{[0]}\cdot F_4^{[2]}}(N)&=\sum_I\left[-\frac{2\mu}{\tau_0R^8}+\frac{8\mu^2}{f_0R^7}\right]\widehat\beta_I(N)\nonumber\\
&=\frac{\mu f_0^2}{25\tau_0}(-18+6)\sum_I\widehat\beta_I(N)\nonumber\\
&=-\frac{12\mu f_0^2}{25\tau_0}\sum_I\widehat\beta_I(N)\,.
\label{eq:F4contractionsteps}
\end{align}
The second line uses $R^4=5/(3f_0)$ and $\tau_0=f_0^2R^3/(20\mu)$. Thus the signed coefficient includes both equations of the four-form system.

For the NSNS form, the diagonal metric variation in~\eqref{eq:first-metric} gives
\begin{equation}
\star_1[g^{[1]}]H_3^{[0]}=\frac{\mu}{2\tau_0}\star_0\sum_I\beta_I\left(\frac{32\mu}{f_0}q_I\alpha_I-H_3^{[0]}\right)\,.
\label{eq:Hstarvariationexplicit}
\end{equation}
Indeed, $\sum_{m\in U_J}\varsigma_{I,m}$ equals $-3$ for $I=J$ and $1$ otherwise. Applying~\eqref{eq:Hodgecontractionidentity} in~\eqref{eq:Hsolution} now gives
\begin{align}
\widehat{H_3^{[0]}\cdot H_3^{[2]}}(N)&=\sum_I\Bigg[\frac{4\mu f_0^2}{25\tau_0^3}-\frac{4\mu f_0^2}{25\tau_0^3}\nonumber\\
&\qquad+\frac{8\mu^2}{\tau_0^2R^3}+\frac{8\mu^2}{f_0\tau_0^2R^7}\Bigg]\widehat\beta_I(N)\nonumber\\
&=\frac{\mu f_0^2}{25\tau_0^3}(4-4+10+6)\sum_I\widehat\beta_I(N)\nonumber\\
&=\frac{16\mu f_0^2}{25\tau_0^3}\sum_I\widehat\beta_I(N)\,.
\label{eq:Hcontractionsteps}
\end{align}
The four terms come, in order, from the scalar prefactor $-\mu\sum_I\beta_I/(2\tau_0)$, the Hodge variation~\eqref{eq:Hstarvariationexplicit}, $f_0\star_0F_2^{[1]}$ and $\star_0F_4^{[0]}\wedge F_2^{[1]}$ in~\eqref{eq:Hsecond}. The first two contributions cancel. The two RR terms leave the positive coefficient above. Linearity adds the four source families, including modes shared by two families. The zero modes of both contractions vanish because the form corrections have zero harmonic part and $\widehat\beta_I(0)=0$. Inverse Fourier transformation gives~\eqref{eq:signedcontractions}.

\section{Expanded Einstein equation and verification}
\label{app:coefficients}
\subsection{Unreduced tensor coefficient}
All raised internal indices in the coefficient below use the zeroth-order inverse metric $g_{[0]}^{mn}=R^{-2}\delta^{mn}$. For example, $H_n^{[2]\,ab}=g_{[0]}^{ac}g_{[0]}^{bd}H^{[2]}_{ncd}$. The inverse-metric corrections appear in the explicitly displayed terms proportional to $g_{[1]}^{ab}$, defined above~\eqref{eq:connectionlinear}. The coefficient multiplying $n^{-1/2}$ in the full internal equation is
\begin{align}
\mathcal E^{[2]}_{mn}={}&\tau_0^2R^{[2]}_{mn}+2\tau_0\tau^{[1]}R^{[1]}_{mn}-2\tau_0\partial_m\partial_n\tau^{[2]}-\frac{4\tau_0^2}{w_0}\partial_m\partial_nw^{[2]}\nonumber\\
&+2\partial_m\tau^{[1]}\partial_n\tau^{[1]}-2\tau^{[1]}\partial_m\partial_n\tau^{[1]}+2\tau_0\Gamma^{[1]p}{}_{mn}\partial_p\tau^{[1]}\nonumber\\
&-\frac{4\tau_0^2}{w_0}\left(\frac{2\tau^{[1]}}{\tau_0}-\frac{w^{[1]}}{w_0}\right)\partial_m\partial_nw^{[1]}+\frac{4\tau_0^2}{w_0}\Gamma^{[1]p}{}_{mn}\partial_pw^{[1]}\nonumber\\
&-\tau_0\tau^{[1]}|H_3^{[0]}|^2_{mn}-\frac{\tau_0^2}{4}\left(H^{[0]}_{mab}H_n^{[2]\,ab}+H^{[2]}_{mab}H_n^{[0]\,ab}\right)\nonumber\\
&+\frac{\tau_0^2}{2}g_{[1]}^{ab}H^{[0]}_{mac}H^{[0]}_{nb}{}^c-\frac1{12}\left(F^{[0]}_{mabc}F_n^{[2]\,abc}+F^{[2]}_{mabc}F_n^{[0]\,abc}\right)\nonumber\\
&+\frac14g_{[1]}^{ab}F^{[0]}_{macd}F^{[0]}_{nb}{}^{cd}-\frac12|F_2^{[1]}|^2_{mn}+\frac14g^{[0]}_{mn}|F_2^{[1]}|^2\nonumber\\
&+\frac12g^{[0]}_{mn}F_4^{[0]}\cdot F_4^{[2]}-\frac14g^{[0]}_{mn}g_{[1]}^{ab}|F_4^{[0]}|^2_{ab}+\frac14g^{[1]}_{mn}\left(f_0^2+|F_4^{[0]}|^2\right)\nonumber\\
&-\frac{\mu}{2}\sum_I\left[\tau^{[1]}\left(\Pi^{[0]}_{I,mn}-\frac12g^{[0]}_{mn}\right)\rho_I^{[0]}+\tau_0\left(\Pi^{[1]}_{I,mn}-\frac12g^{[1]}_{mn}\right)\rho_I^{[0]}\right]\nonumber\\
&-\frac{\mu\tau_0}{2}\sum_I\left(\Pi^{[0]}_{I,mn}-\frac12g^{[0]}_{mn}\right)\rho_I^{[1]}=0\,.
\label{eq:einsteinsecond}
\end{align}
The symbols $H$ and $F$ in the component contractions denote the three-form and four-form. For the diagonal first-order metric, $\Pi^{[1]}_{I,mm}=g^{[1]}_{mm}$ in a tangent direction and vanishes in a transverse direction. The source-density variation is given in equation~\eqref{eq:sourcedensityvariation}.

\subsection{Metric verification}
\label{app:metricsolution}
The substitution uses the unreduced Ricci operator~\eqref{eq:Riccisymbol}, rather than only its gauge-reduced form. At positive heat time we retain $\rho_I^{[0]}=8/R^3-\Delta_0\beta_I$, with the heat label suppressed in this appendix.

For one family, combining the scalar Hessians with all remaining quadratic terms in~\eqref{eq:einsteinsecond} gives
\begin{equation}
\left.\mathcal E^{[2]}_{mn}\right|_{g^{[2]}=0,\,\beta_I\beta_I}=\frac{\mu^2}{2}\partial_m\partial_n\beta_I^2+\frac{\mu^2}{4}\left(\frac34\Pi^{[0]}_{I,mn}-\frac14g^{[0]}_{\perp I,mn}\right)\Delta_0\beta_I^2\,.
\label{eq:selftensorsource}
\end{equation}
It is canceled by the unreduced Ricci tensor of
\begin{align}
g^{[2]}_{mn}\big|_{\rm self}=\frac{\mu^2}{2\tau_0^2}\sum_I\Bigg[&\left(\frac34\Pi^{[0]}_{I,mn}-\frac14g^{[0]}_{\perp I,mn}\right)\nonumber\\
&\times\left(\beta_I^2-\int_{T^6}\dd^6u^{\prime}\,\beta_I^2(u^{\prime})\right)\nonumber\\
&+2\partial_m\partial_n\Delta_0^{-1}\left(\beta_I^2-\int_{T^6}\dd^6u^{\prime}\,\beta_I^2(u^{\prime})\right)\Bigg]\,.
\label{eq:metricsingular}
\end{align}
The Hessian in this expression has zero linearized Ricci tensor, but restores the coefficient gauge: the gauge divergence of the diagonal term is canceled by that of the Hessian. Its Fourier transform is~\eqref{eq:metricselfexplicit}. The complete self-source equation is therefore a derivative of an integrable square, even though its separated terms are singular.

For two different families, give their profiles arbitrary wave covectors supported in their respective transverse triples. Substituting~\eqref{eq:mixedscalarderivative} into the scalar Hessians and using~\eqref{eq:Riccisymbol}, clearing the fourth power of the total wavevector norm makes the equation a polynomial identity. The dependence on the split of the common transverse momentum cancels, leaving~\eqref{eq:metricmixedexplicit}. The linear-flux identity follows in the same way from the form coefficients. This establishes the nonzero-mode equations at positive heat time. The limit follows from~\eqref{eq:productlimit}.

For the tensor mean, integrate $\beta_I\partial_a\partial_b\beta_J$ by parts at positive heat time. Opposite momenta can belong to both transverse lattices only along their common coordinate $c$, so a nonzero derivative-product mean requires $a=b=c$. The coefficients of all these surviving products cancel in each of the 21 independent tensor components. The self-source means vanish by~\eqref{eq:selftensorsource}, and the linear means vanish by $\int_{T^6}\dd^6u\,\beta_I=0$. This proves zero-mode solvability before the regulator is removed, without assigning independent finite parts to divergent self-source terms.

\subsection{Integrated equations for the first-order constants}
\label{app:homogeneous}
The first-order differential equations allow constants $C_w^{[1]}$, $C_\tau^{[1]}$ and $C_{g,mm}^{[1]}$ in the warp factor, dilaton and six diagonal metric entries. Their values are determined by the next-order zero modes. In the common-heat prescription, the nonconstant quadratic sources have vanishing means as shown above. With zero harmonic form corrections, the remaining mean equations are obtained by varying the metric and scalar constants in the algebraic smeared equations while keeping the flux forms fixed in the coordinate basis, as in section~\ref{sec:consistency}. The two scalar equations reduce to
\begin{align}
0&=8\frac{C_\tau^{[1]}}{\tau_0}-3\frac{C_w^{[1]}}{w_0}+\sum_m\frac{C_{g,mm}^{[1]}}{R^2}\,,\\
0&=46\frac{C_\tau^{[1]}}{\tau_0}-7\sum_m\frac{C_{g,mm}^{[1]}}{R^2}\,.
\label{eq:homogeneous-scalars}
\end{align}
Let $\bar m$ be the partner of $m$ on the same two-torus. The six diagonal Einstein equations are
\begin{equation}
0=-8\frac{C_\tau^{[1]}}{\tau_0}+6\frac{C_{g,mm}^{[1]}}{R^2}+7\frac{C_{g,\bar m\bar m}^{[1]}}{R^2}+2\sum_{a\notin\{m,\bar m\}}\frac{C_{g,aa}^{[1]}}{R^2}\,.
\label{eq:homogeneous-tensor}
\end{equation}
Subtracting the equations for $m$ and $\bar m$ equates the two shifts in each pair; subtracting equations for distinct pairs equates all six. Their common value obeys $21C_{g,mm}^{[1]}/R^2=8C_\tau^{[1]}/\tau_0$. Inserting this relation into the dilaton constraint gives $30C_\tau^{[1]}/\tau_0=0$. The metric shifts and then the warp-factor shift vanish. The unequal coefficients six and seven include the variation of the transverse source density; dropping it would change the system.

Thus the system has rank eight and gives~\eqref{eq:homogeneouszero}. These constraints use the common heat limit and the zero harmonic form corrections specified above.

\bibliographystyle{JHEP}
\bibliography{2ndorderT6Z2Z2}
\end{document}